\PassOptionsToPackage{usenames,dvipsnames,svgnames,table}{xcolor}
\documentclass[10pt,journal,compsoc]{IEEEtran} 

\usepackage{textcomp} \usepackage[pdftex]{graphicx}
\usepackage[utf8]{inputenc}
\usepackage[T1]{fontenc}
\usepackage{fontawesome}
\usepackage{url}
\usepackage{xspace}
\usepackage{xcolor}
\usepackage{marvosym}
\usepackage[export]{adjustbox}
\usepackage{booktabs, tabularx, multirow, makecell, siunitx}
\usepackage{eso-pic}
\usepackage[most]{tcolorbox}
\usepackage{calc}
\usepackage{ragged2e}

\usepackage{siunitx}
\usepackage[style=ieee,minnames=3,maxnames=3]{biblatex}
\AtBeginBibliography{\footnotesize}
\AtEveryBibitem{\clearname{editor}\clearlist{location}\clearfield{series}\clearfield{pagetotal}\clearfield{pages}\clearfield{doi}\clearfield{isbn}\clearfield{issn}\clearfield{booksubtitle}\ifentrytype{online}{}{\ifentrytype{report}{}{\ifentrytype{article}{}{\clearfield{url}}}}\ifentrytype{online}{}{\clearfield{month}}\ifentrytype{inproceedings}{\clearfield{year}}{}}
\newcolumntype{R}{>{\raggedleft\arraybackslash}X}

\usepackage{setspace}

\usepackage{acronym}
\usepackage{amsmath,amssymb,amsfonts}
\usepackage[abbreviations]{foreign}
\usepackage[colorlinks=true,linkcolor=black,anchorcolor=black,citecolor=black,filecolor=black,menucolor=black,runcolor=black,urlcolor=black]{hyperref}
\usepackage[capitalise]{cleveref}
\usepackage{pifont}
\usepackage[caption=false]{subfig}
\usepackage{orcidlink}

\definecolor{redindiagram}{HTML}{B1001C}

\newcommand{\twine}{\textsc{Twine}\xspace}
\newcommand{\sys}{\twine}
\newcommand{\acctee}{\textsc{AccTEE}\xspace}
\newcommand{\polybench}{PolyBench/C\xspace}
\newcommand{\ipfs}{Intel protected file system\xspace}

\newcommand{\xm}{\emph{Credora}\xspace}
\newcommand{\xmi}{\emph{Credora Inc.}\xspace}
\newcommand{\Ding}[1]{\raisebox{-0.8pt}{\ding{\the\numexpr #1 + 191}}}
\newcommand{\DingBlack}[1]{\raisebox{-0.8pt}{\ding{\the\numexpr #1 + 201}}}
\newcommand{\DingRed}[1]{\raisebox{-0.8pt}{\color{redindiagram}\ding{\the\numexpr #1 + 201}}}

\newcommand*{\paratitle}[1]{\vspace{3pt}\noindent\textbf{#1.}\enspace}

\DeclareSIUnit\tps{TPS} \DeclareSIUnit{\nothing}{\relax}

\usepackage[inline]{enumitem}
\setlist{noitemsep,topsep=0pt,parsep=0pt,partopsep=0pt,leftmargin=3mm}
\def\insertLklMemBepc{2.9}
\def\insertLklFileBepc{0.9}
\def\insertLklMemAepc{16.3}
\def\insertLklFileAepc{1.0}
\def\insertWamrMem{1.9}
\def\insertWamrFile{1.1}
\def\insertTwineMemBepc{4.3}
\def\insertTwineFileBepc{2.3}
\def\insertTwineMemAepc{10.4}
\def\insertTwineFileAepc{2.4}
\def\insertLklVsTwineMemAepc{1.3}

\def\readSeqLklMemBepc{2.0}
\def\readSeqLklFileBepc{1.3}
\def\readSeqLklMemAepc{7.7}
\def\readSeqLklFileAepc{3.1}
\def\readSeqWamrMem{1.4}
\def\readSeqWamrFile{1.3}
\def\readSeqTwineMemBepc{3.1}
\def\readSeqTwineFileBepc{3.8}
\def\readSeqTwineMemAepc{6.3}
\def\readSeqTwineFileAepc{8.3}

\def\readRandLklMemBepc{2.3}
\def\readRandLklFileBepc{20.4}
\def\readRandLklMemAepc{17.6}
\def\readRandLklFileAepc{22.7}
\def\readRandWamrMem{1.3}
\def\readRandWamrFile{1.3}
\def\readRandTwineMemBepc{2.9}
\def\readRandTwineFileBepc{18.3}
\def\readRandTwineMemAepc{17.6}
\def\readRandTwineFileAepc{20.1}

\def\readRandLklVsTwineFileBepc{1.1}
\def\readRandLklVsTwineFileAepc{1.13}
 \def\cryptoSymRatioWasmToNative{1.2}
\def\cryptoSymRatioWasmSgxToNative{1.3}
\def\cryptoHashRatioWasmToNative{1.3}
\def\cryptoHashRatioWasmSgxToNative{1.4}
\def\cryptoAsymRatioWasmToNative{3.3}
\def\cryptoAsymRatioWasmSgxToNative{5.1}
 \def\tlsAverageNative{157}
\def\tlsAverageSgx{97}
\def\tlsAverageWasm{67}
\def\tlsAverageTwine{44}
\def\tlsRatioSgx{1.6}
\def\tlsRatioWasm{2.4}
\def\tlsRatioTwine{3.6}
\def\tlsRatioSgxVsTwine{2.2}
 \def\speedtestWamrMemToNativeRatio{1.7}
\def\speedtestWamrFileToNativeRatio{1.6}
\def\speedtestTwineMemToWamrRatio{2.1}
\def\speedtestTwineFileToWamrRatio{2.0}
\def\speedtestExpFourOneZeroTwineMemVsFileRatio{32.5}
\def\speedtestExpFourOneZeroSgxLklMemVsFileRatio{22.0}
\def\speedtestTwineMemRatioUpperPlot{3.9}
\def\speedtestTwineFileRatioUpperPlot{4.6}
 \def\compilationNativeAverage{23,350}
\def\compilationSgxAverage{288,774}
\def\compilationWasmBinAverage{4,329}
\def\compilationWasmSgxBinAverage{3,425}
\def\compilationWasmAverage{38,593}
\def\optimizationSgxLklDiskImageAverage{15,711}
\def\optimizationWasmAotAverage{52,944}
\def\launchTimeNativeAverage{2}
\def\launchTimeSgxAverage{6,119}
\def\launchTimeWasmAverage{70}
\def\launchTimeWasmSgxAverage{3,155}
\def\memoryNativeInMemoryAverage{192,822}
\def\memorySgxInMemoryAverage{77,310}
\def\memoryWasmInMemoryAverage{211,156}
\def\memoryWasmSgxInMemoryAverage{9,970}
\def\memorySgxEnclaveSize{261,120}
\def\memoryWasmSgxEnclaveSize{209,920}
\def\sizeNativeAverage{1,164}
\def\sizeSgxLklBinAverage{6,546}
\def\sizeWasmBinAverage{123}
\def\sizeWasmSgxBinAverage{30}
\def\sizeSgxLklEnclaveAverage{79,200}
\def\sizeWasmSgxEnclaveAverage{567}
\def\sizeSgxLklDiskImageAverage{247,552}
\def\sizeWasmAverage{1,155}
\def\sizeWasmAotAverage{3,707}
\def\launchTimeTwineVsSgx{1.9}
 \def\memsetRatio{50.1}
\def\ocallRatio{36.2}
\def\otherOperationsRatio{10.7}
\def\sqliteRatio{2.9}
\def\copyRatio{75.9}
\def\copyInvertedRatio{24.1}
\def\insertRatio{1.5}
\def\seqReadingRatio{2.5}
\def\randReadingRatio{4.1}
 
\makeatletter
\begin{document}

\title{A Comprehensive Trusted Runtime for WebAssembly with Intel SGX}

\author{Jämes~Ménétrey\,\orcidlink{0000-0003-2470-2827},~\IEEEmembership{Graduate Student Member,~IEEE,}
        Marcelo~Pasin\,\orcidlink{0000-0002-3064-5315},~\IEEEmembership{Member,~IEEE},\\
		Pascal~Felber\,\orcidlink{0000-0003-1574-6721},~\IEEEmembership{Senior Member,~IEEE,}
        Valerio~Schiavoni\,\orcidlink{0000-0003-1493-6603},~\IEEEmembership{Member,~IEEE},
		Giovanni~Mazzeo\,\orcidlink{0000-0002-0238-5616},~\IEEEmembership{Member,~IEEE},
		Arne~Hollum\,\orcidlink{0009-0005-0688-3953},
		Darshan~Vaydia\,\orcidlink{0009-0006-4977-1852}
		
\IEEEcompsocitemizethanks{
\IEEEcompsocthanksitem Jämes Ménétrey (corresponding author), Marcelo Pasin, Pascal Felber and Valerio Schiavoni are with the University of Neuchâtel, Switzerland\\
(e-mail: \href{mailto:james.menetrey@unine.ch}{james.menetrey@unine.ch}; \href{mailto:firstname.lastname@unine.ch}{firstname.lastname@unine.ch}).
\IEEEcompsocthanksitem Giovanni Mazzeo is with the University of Naples 'Parthenope', Italy 
(e-mail: \href{mailto:giovanni.mazzeo@uniparthenope.it}{giovanni.mazzeo@uniparthenope.it}).
\IEEEcompsocthanksitem Arne Hollum and Darshan Vaydia are with Credora Inc., US\\
(e-mail: \href{mailto:arne@credora.io}{arne@credora.io}; \href{mailto:darshan@credora.io}{darshan@credora.io}).
}}

\markboth{{IEEE} Transactions on Dependable and Secure Computing}{Ménétrey \MakeLowercase{\textit{et al.}}: \twine: An Embedded Trusted Runtime for WebAssembly (to change)}

\IEEEtitleabstractindextext{

\begin{abstract}
\justifying
In real-world scenarios, trusted execution environments (TEEs) frequently host applications that lack the trust of the infrastructure provider, as well as data owners who have specifically outsourced their data for remote processing.
We present \twine, a trusted runtime for running WebAssembly-compiled applications within TEEs, establishing a two-way sandbox.
\twine leverages memory safety guarantees of WebAssembly (Wasm) and abstracts the complexity of TEEs, empowering the execution of legacy and language-agnostic applications. 
It extends the standard WebAssembly system interface (WASI), providing controlled OS services, focusing on I/O.
Additionally, through built-in TEE mechanisms, \twine delivers attestation capabilities to ensure the integrity of the runtime and the OS services supplied to the application.
We evaluate its performance using general-purpose benchmarks and real-world applications, showing it compares on par with state-of-the-art solutions.
A case study involving fintech company \xm reveals that \twine can be deployed in production with reasonable performance trade-offs, ranging from a 0.7$\times$ slowdown to a 1.17$\times$ speedup compared to native run time.
Finally, we identify performance improvement through library optimisation, showcasing one such adjustment that leads up to \randReadingRatio$\times$ speedup.
\twine is open-source and has been upstreamed into the original Wasm runtime, WAMR.
\end{abstract}
 \begin{IEEEkeywords}
WebAssembly, trusted execution environment, remote attestation, protected file system, protected database, Intel SGX
\end{IEEEkeywords}
}

\maketitle
\IEEEdisplaynontitleabstractindextext

\def\confname{IEEE Transactions on Dependable and Secure Computing (TDSC)}
\def\confyear{2023}
\def\confdoi{10.1109/TDSC.2023.3334516}

\definecolor{yellowPaper}{HTML}{fff8ae}
\AddToShipoutPictureFG*{\AtTextUpperLeft{\raisebox{-10pt}{\hspace*{-7pt}
        \begin{tcolorbox}[width=\dimexpr(\textwidth + 10pt),colback=yellowPaper,enhanced,frame hidden,sharp corners]  
            \centering\scriptsize
            \copyright~\confyear\ IEEE. Personal use of this material is permitted. Permission from IEEE must be obtained for all other uses, in any current or future media, including reprinting/republishing this material for advertising or promotional purposes, creating new collective works, for resale or redistribution to servers or lists, or reuse of any copyrighted component of this work in other works.
            This is the author's version of the work. The definitive version has been published  in the journal of the 
            \confname.
            \href{https://doi.org/\confdoi}{DOI: \confdoi}
        \end{tcolorbox}
    }}} 

\IEEEraisesectionheading{\section{Introduction}\label{sec:intro}}

\IEEEPARstart{D}{ata} confidentiality and secure code execution are fundamental components for many organisations in the current data-driven and interconnected landscape. 
Trusted execution environments (TEEs) such as Intel SGX/TDX~\cite{cryptoeprint:2016:086,Intel2020TDX}, Arm TrustZone/CCA~\cite{pinto2019demystifying,Arm2021CCA}, AMD SEV/SEV-SNP~\cite{amd-sev,sev2020strengthening} and RISC-V TEEs~\cite{keystone,sanctum,timberv,garlati2020clean} gathered much attention lately as they provide hardware support for secure code execution within special hardware constructs (\ie enclaves) shielded from the outside world, including a malicious or compromised operating system and privileged users.
In a typical TEE usage model, there are multiple entities that do not always trust each other: \emph{i}) the \emph{TEE service provider} that provides the enclave code for the confidential processing; \emph{ii}) the \emph{infrastructure provider} hosting the TEE-based service; \emph{iii}) and the \emph{data owner} who outsourced its data for remote processing. 
The TEE technology gives security guarantees to service providers and data owners against a malicious infrastructure provider. Further, there are no consolidated approaches for ensuring trust to data owners or infrastructure providers that completely mistrust the TEE service provider. This is even more problematic when the enclave program itself is private and cannot be exposed. Programs may have exploitable bugs or write information out of the enclave through corrupted or dangling pointers.

To cope with the situation explained, researchers explored two distinct avenues: leveraging methodologies of Software Fault Isolation (SFI) or taking advantage of WebAssembly (Wasm) inside TEEs.
SFI~\cite{sfi} establishes a logical protection domain by inserting dynamic checks before memory and control-transfer instructions, which are then verified at runtime.
Wasm~\cite{10.1145/3140587.3062363}, initially conceived for executing high-performance native code in browsers, allows building a memory-safe, lightweight, and portable sandboxed execution environment based on restricted memory access and control flow with limited usage of resources.
In the field of SFI, \emph{Ryoan}~\cite{ryoan2018TOCS} and \emph{Deflection}~\cite{deflection} are notable solutions which combined SFI with TEEs.
However, both suffer from performance and usability issues. In fact, these runtimes do not allow the execution of unmodified source code.
In the domain of Wasm, we highlight \acctee~\cite{DBLP:journals/corr/abs-1908-11143} and SGX-LKL~\cite{DBLP:conf/middleware/GoltzscheNKK19}, which execute Wasm binaries inside secure enclaves.
\acctee provides resource accounting for Wasm bytecode but lacks comprehensive OS-level sandboxing, relying instead on SGX-LKL for enclave execution. On the other hand, SGX-LKL serves as a library OS designed for generic code execution but does not offer sandboxing capabilities and exposes a large trusted computing base (TCB) due to its integration with the Linux kernel library.

In this paper, we extend our previous work~\cite{menetrey2021twine}, and present \sys, a runtime that supports the execution of WebAssembly-compilable software --- \ie potentially legacy --- in trusted execution environments, providing capabilities of a two-way sandbox. \twine comes with an extended WebAssembly system interface (WASI), which allows the sandboxed application to issue filtered and controlled OS services. 
We currently support Intel SGX as a TEE: \twine translates at runtime WASI operations into equivalent native OS calls or functions in secure libraries specifically built for Intel SGX.
In particular, \twine maps all file operations to \ipfs~\cite{ipfs}, and persisted data is transparently encrypted and never accessible in plaintext from outside an enclave, thus shielding against data exfiltration. Furthermore, our solution provides configurable communication capabilities to the Wasm binary running inside the TEE using raw TLS or HTTPS. Last but not least, \sys{} includes attestation features, which allow us to verify the integrity of the entire runtime and, most importantly, the way OS services are offered to the Wasm module.  
While a TEE provides a secure hardware execution runtime in the processor, \twine provides a secure software runtime (sandbox) nested within the TEE, using WASI for interoperability with regular API and abstracting the underlying environment from the application.

We evaluated \twine with several micro- and macro-benchmarks.
We compared the performances of \twine against existing alternatives, with and without secure operations inside a TEE.
Our results reveal that \twine performs on par with systems also exploiting TEEs and providing similar security guarantees while offering programmers greater freedom.
We observed performance overheads for some specific workloads due to execution within the TEE.
In addition, we integrated and deployed in production the resulting trustworthy runtime in the context of a cryptocoin-based fintech company, which, in short, computes real-time credit scoring, achieving 0.7$\times$--1.17$\times$ the performance of their original native software.
We believe this penalty is largely compensated by the additional security guarantees and the interoperability granted by our trusted WASI layer.

This publication encompasses significant revisions compared to the initial prototype previously published~\cite{menetrey2021twine}.
Technical advancements include the integration of secure communication (\S\ref{sec:network}), remote attestation capabilities (\S\ref{sec:ra}), and a collaboration with a fintech company, \xm, illustrating practical application (\S\ref{sec:usecases}).
Additionally, this manuscript presents updated evaluations with new benchmarks for cryptographic primitives and communication throughput (\S\ref{sec:networkstack}), an evaluation of \sys in the production environment of \xm for crypto-coins credit scoring (\S\ref{sec:xm}), and refreshed results from previous benchmarks (\S\ref{sec:polybench}, \S\ref{sec:speedtest1}) alongside an analysis of large memory pages' impact.

\noindent The contributions presented in this paper are:
\begin{itemize}
    \item The first fully open-source implementation of a general-purpose Wasm runtime environment within SGX enclaves supporting encrypted file system and networking, as well as built-in abilities for remote attestation;
    \item An extensive experimental evaluation, shedding light on performance costs and associated bottlenecks, as well as a real-world integration in an industry-battled scenario;
    \item A proposal for improving Intel protected file system and a showcase of the derived performance improvements.
    \item A fully upstreamed solution to the original Wasm runtime, which is open-source and ready for use in production environments.
\end{itemize}
This paper is organised as follows. 
We provide a background on Intel SGX and WebAssembly in \S\ref{sec:backg}.
In \S\ref{sec:relat}, we survey related work.
We detail the design and implementation of \twine in \S\ref{sec:design} and \S\ref{sec:runti}, respectively.
We describe the credit-scoring scenario in \S\ref{sec:usecases}.
We report on the evaluation of our full prototype in \S\ref{sec:evalu}.
The security aspects of our approach are examined in \S\ref{sec:secu}.
Finally, we conclude in \S\ref{sec:concl}.

\section{Background}\label{sec:backg}

This section provides background information on Intel SGX in (\S\ref{sec:sgx}) and the Wasm ecosystem (\S\ref{sec:wasm}) to help understand the architecture and design of \sys.

\subsection{Intel SGX}\label{sec:sgx}

Software Guard Extensions (SGX)~\cite{cryptoeprint:2016:086} are extensions to Intel's instruction set available in modern server-grade processors.
These extensions let developers create encrypted memory regions, \ie \emph{enclaves}.
Enclave memory content is automatically encrypted and decrypted by a dedicated hardware co-processor when accessed (\eg read, write) by instructions running inside the enclave.
Enclave encryption keys are kept inside the processor, and no instruction gives access to them, not even when running with high hardware privilege levels, as OSs and virtual machine managers do.
The memory inside an enclave is protected from unauthorised access, even from attackers with physical access.

Enclave memory access is accelerated by the enclave page cache (EPC).
Intel SGX comes in two editions: \emph{1)} the legacy \emph{Client} SGX with an EPC size up to 256 MiB, and \emph{Scalable} SGX, with latest CPUs, with an EPC size up to 512 GiB.
Client SGX maintains cryptographic hashes for all enclave pages in the EPC so that a modification from outside an enclave can be automatically detected (\ie using a Merkle tree).
On the opposite, Scalable SGX relies on Intel TME~\cite{intel2021tme} to protect the EPC, offering a larger page cache at the cost of a reduced set of hardware attack mitigations~\cite{johnson2021supporting}.
The EPC helps reduce access time to encrypted memory and limits the number of pages concurrently available.
Swapping degrades performance, and enclaved applications should strive to avoid it~\cite{10.1145/3064176.3064219}.

Instructions inside enclaves can access data outside the enclave: to do so, one relies on special \emph{exit call} instructions (\texttt{OCALL}).
Upon an \texttt{OCALL}, the CPU exits the protected enclave to execute outside (unsafe) code.
Conversely, an \emph{enter call} (\texttt{ECALL}) instruction is required to call code inside an enclave.
\texttt{OCALL} and \texttt{ECALL} instructions are slow because switching the context (\eg from inside to outside an enclave, and vice versa) is costly, with a median of 14'170 CPU cycles~\cite{weisse2017regaining}.
Applications using enclaves can reduce the performance loss of context switches using workarounds~\cite{10.1145/3268935.3268942} or optimisations~\cite{yuhala2023switchless}.

Building distributed and complex software artefacts using enclaves requires establishing trust across the different components.
Intel SGX natively supports remote attestation (RA) for that purpose.
Specifically, RA can be used to ensure the trustworthiness of the enclave code (\eg nobody tampered with the executable).
Each processor has a secret key fused in its die, used to derive many keys.
One of the derived keys is used to build enclave attestations, calculated as a signature of the entire content of an enclave at its creation (\ie a \emph{measurement}).
As a result, an external attestation service confirms that a given enclave runs a particular piece of code on a genuine Intel SGX processor.
\vspace*{35pt}
\pagebreak

\subsection{WebAssembly}\label{sec:wasm}

WebAssembly (Wasm) is a W3C open standard for a portable and executable bytecode format.
It was initially designed to improve the performance of Web applications with near-native speed and has since been supported in standalone environments (\ie outside browsers).
Full application execution, especially in standalone environments, requires access to OS services, \eg process and memory management or I/O, typically available via standard system calls (for instance, exposed by a POSIX interface).
Hence, the interaction of Wasm with the underlying OS is standardised through a specific API called WebAssembly system interface (WASI)~\cite{Mozilla2019StandardizingWASI}.
This interface allows for several implementations suited to different OSs and incorporates several non-functional abstractions, such as virtualisation, sandboxing and access control.
In the latest specifications, the WASI interface consists of a handful of functions covering various system capabilities such as file system, network and time access, events polling and random number generation.
Wasm has been advocated as an ideal homogeneous environment that could be used to create cross-platform applications spanning cloud, edge, and IoT devices, known as the \emph{cloud-edge continuum}~\cite{DBLP:conf/hpdc/MenetreyPFS22}.
\sys is an example of a trusted runtime that can be deployed in the cloud.

WebAssembly binaries are generated by either using the \emph{Emscripten}~\cite{10.1145/2048147.2048224} or the LLVM~\cite{DBLP:conf/cgo/LattnerA04} compiler for browser- and server-oriented execution, respectively. The execution of Wasm must be handled by a dedicated runtime, able to execute the instructions and implement WASI calls.
Prior work extensively analysed the state-of-the-art Wasm runtimes~\cite{menetrey2021twine,wang2022compareruntimes}.
We settled for the WebAssembly Micro Runtime (WAMR)~\cite{wamr} from the Bytecode Alliance for many reasons.
First, WAMR supports Intel SGX off the shelf and is written in C, a language supported by the SDK of SGX.
Second, WAMR offers many execution modes, amongst which is the ahead-of-time compiler, leading to near-native speed performance.
Finally, the runtime has a small footprint because of its lightweight implementation and the lack of dependency on external programming libraries, reducing the size of the TCB.

WAMR can execute code in three modes, each with its benefits and drawbacks: interpreted, compiled just-in-time (JIT) and compiled ahead-of-time (AOT).
The interpreted mode has two strategies: the classic interpreter, a textbook implementation of a Wasm interpreter, and the fast interpreter, which precompiles the Wasm opcode into an internal representation, yielding a performance increase of 2$\times$.
JIT mode also has two strategies: the LLVM JIT, based on the LLVM framework and offers the best run time performance, and the fast JIT, a handcrafted lightweight JIT engine with a small footprint.
Finally, AOT mode achieves a near-native speed, has a small footprint and a fast startup time.
In the context of Intel SGX, achieving the fastest execution while minimising the attack surface of the TCB is of utmost importance.
For these reasons, we opted for using the AOT execution mode for \sys.
We believe this approach does not remove the versatile and portable characteristics of Wasm since the final binary artefacts are generated from the Wasm bytecode~\cite{bytecodealliance2023wamrmodes}.
 \section{Related Work}\label{sec:relat}

We survey related work based on different criteria.
First, we look at systems with dedicated support for Wasm in TEEs.
Then, we review proposals for generic TEE support for language runtimes.
Finally, we investigate alternative sandboxing solutions embedded in TEEs.

\paratitle{WebAssembly and TEEs}
\acctee~\cite{DBLP:conf/middleware/GoltzscheNKK19} runs Wasm inside Intel SGX, with the specific goal of implementing trustworthy resource accounting under malicious OSs.
It leverages the SGX-LKL~\cite{DBLP:journals/corr/abs-1908-11143} library OS and V8 JavaScript/WebAssembly engine to execute Wasm binaries inside SGX enclaves. 
Their two-way sandbox (firstly from disjoint memory spaces for Wasm modules, and secondly from SGX itself) is similar to \twine's double-sandboxing approach (see \S\ref{sec:runti} for details).
\acctee lacks a WASI layer and instead uses custom JavaScript for I/O interfacing, managed by SGX-LKL.
This absence of a WASI layer presents two issues: firstly, a performance slowdown due to the constant call of JavaScript for external interactions, and secondly, a sandboxing challenge since \acctee uses Emscripten for Wasm compilation, which can trigger system calls.
In contrast, \twine includes WASI to securely and efficiently handle data persistence and communication

Se-Lambda~\cite{10.1007/978-3-030-01701-9_25} is a library built on top of OpenLambda~\cite{hendrickson2016serverless} to deploy Wasm-based serverless programs over Function-as-a-Service (FaaS) platforms.
Se-Lambda shields the FaaS gateway and the code of the deployed functions inside enclaves, providing anti-tampering and integrity guarantees.
Further, it protects against attacks from a privileged monitoring module that intercepts and checks system call return values.
We believe that similar defence mechanisms could be easily integrated into \sys.

Enarx~\cite{enarx} represents an open-source TEE runtime that facilitates the execution of Wasm binaries across various TEEs, including those compatible with Intel and AMD processors.
Although both Enarx and \sys extend support for network and attestation capabilities, there are notable distinctions between the two.
Specifically, Enarx lacks file system support and solely relies on a just-in-time compiler.
In comparison, \sys provides a robust safeguard for data at rest, as well as an ahead-of-time compiler that ensures fast start-up and execution.

Veracruz~\cite{veracruz} protects workloads running on Arm CCA and AWS Nitro~\cite{nitro}.
Unlike \sys, Veracruz is a framework for developing Wasm applications in a confidential cloud computing context rather than an arbitrary execution environment for unmodified software.

\textsc{WaTZ}~\cite{menetrey2022watz} executes ahead-of-time compiled Wasm applications in Arm TrustZone.
\textsc{WaTZ} implements remote attestation for TrustZone.
However, it does not provide a WASI layer for file system and networking capabilities.
\sys reuses the built-in capabilities of SGX's RA, while exposing an easy-to-use API for the hosted applications.

\paratitle{Embedding language runtimes in TEEs}
There have been many efforts to embed other language runtimes into TEEs~\cite{santos2014armnet,mesapy,tsai2020civet,goltzsche2017trustjs,wang2019running}.
\sys deploys a lightweight and versatile Wasm runtime inside an SGX enclave, able to execute AOT-compiled Wasm applications for optimal performance.
Additionally, we developed a WASI layer to enable any WASI-compliant application to run inside our runtime seamlessly.

\paratitle{Sandboxing inside TEEs}
Lastly, we report research works that applied data confinement solutions --- based on Software Fault Isolation --- inside TEEs to protect against untrusted enclave service providers. \emph{Ryoan} \cite{ryoan2018TOCS} introduced a distributed sandbox by adapting the Google Native Client (NaCl) to the enclave environment, thereby containing untrusted data-processing modules to prevent any leakage of user input data. The solution comes with a verifier and a service runtime. The verifier disassembles the binary and validates the disassembled instructions as being safe to execute, to guarantee that the untrusted module cannot break out of NaCl's SFI sandbox.
\\Liu \etal \cite{deflection} presented the \emph{Deflection} model employing out-of-enclave targeted instrumentation for in-enclave information-flow control. Deflection uses 
proof-carrying code (PCC), a technique that enables a verification condition generator to analyze a program and create a proof that attests to the program's adherence to policies, and a proof
checker to verify the proof and the code. Deflection constructs the binary code for a data-processing program and enhances it with security annotations for real-time policy enforcement. Simultaneously, a trusted code consumer operates within the bootstrap enclave to statically verify whether the target code genuinely incorporates the required security annotations.
\\The main drawbacks we highlight in these works are related to performance and usability. Ryoan is particularly intrusive on the applications' binary due to its security annotations. This entails a substantial overhead (\eg 100\% on genetic data). Deflection outperforms Ryoan, with a maximum overhead of 32\%. Both Ryoan and Deflection cannot run legacy code inside the TEE since they require the dedicated partitioning typical of the SGX SDK.

\section{\twine Design}\label{sec:design}
Our objective is to create a runtime for running legacy WebAssembly modules inside a secure enclave, thus enabling a two-way sandbox that ensures typical TEE security features while also mitigating and controlling data leaks, even from malicious TEE service providers.
The runtime must provide dedicated support to the Wasm module, enabling a secure execution of essential OS services within the enclave.

\subsection{Requirements Elicitation}\label{sec:req}
We identify five requirements that guide \twine's design.

\textbf{R1}: \emph{Must support the execution of legacy software services inside the two-way sandbox}. \twine{} should ensure transparent execution of services whose source code can be compiled to WebAssembly. 

\textbf{R2}: \emph{Must provide communication support inside the two-way sandbox}. Wasm modules running inside secure enclaves should be able to communicate through selected application-level protocols or secured socket API.

\textbf{R3}: \emph{Must provide file management support inside the two-way sandbox}. Wasm modules running inside secure enclaves should be able to perform restricted file system operations based on runtime-defined permissions or access paths.

\textbf{R4}: \emph{Must prevent covert channels through the available OS services}. \twine{} should protect the exfiltration of sensitive data from inside the enclave, which may occur through OS service interfaces.

\textbf{R5}: \emph{Must be verifiable}. \twine{} should provide support for verifying the integrity of the runtime and validating its related security policies.

\subsection{Threat model}

\twine leverages the protection of TEEs to offer a trusted environment for running Wasm applications.
Many guarantees offered by \twine are inherited from the underlying TEE, specifically Intel SGX in our implementation.
Note that a different TEE may withstand a different level of threats.

\paratitle{Assumptions}
We assume that adversaries may have physical access to the computer hardware.
However, due to the hardware-intrusion defences offered by Intel SGX, hardware attacks are deemed impractical.
The TEE delivers a level of protection that aligns with Intel's claims, and standard cryptographic techniques cannot be subverted.
Enclave codes present no vulnerabilities by implementation mistake.

\paratitle{SGX enclaves}
Code and data inside enclaves are trusted, while outside entities are considered untrusted.
The non-enclaved part of a process, the OS and any hypervisor are thus potentially hostile.
The memory of an enclave can only be read encrypted from the outside.
Writing the memory enclave from the outside causes the enclave to be terminated.
Side-channel, denial-of-service, fork, or reboot attacks may exist, and applications running inside enclaves must be written to be resistant to them.
While we consider these attacks out of scope, mitigations exist~\cite{10.1145/3359789.3359809,10.5555/3277355.3277378}.

\paratitle{Operating system}
The OS follows an honest-but-curious threat model, and it conforms to its specification,  posing no threat to user and kernel processes.
A compromised OS may arbitrarily respond to enclave calls, causing its malfunction; enclaves should be carefully crafted to ignore abnormal responses or even abandon execution in such cases.

\paratitle{Wasm}
The Wasm specifications and implementation (\eg sandbox, virtual machine) are sane and do not present vulnerabilities.
Side-channel attacks are beyond the scope of this work~\cite{puddu2022the}.

\paratitle{Two-way sandbox}
In the \twine runtime, the threat model for hosted applications is based on an honest-but-curious adversary.
The first layer of defence relies on Wasm and WASI sandboxing to mitigate unauthorised code execution and data modification, shielding the infrastructure owner.
In contrast, the second layer leverages Intel SGX to counter unauthorised data access and code tampering, aimed at securing the application owner.
We further analyse the security implications of AOT-compiled code in \S\ref{sec:secu}.

\begin{figure}[!t]
    \centering
    \includegraphics[scale=0.7]{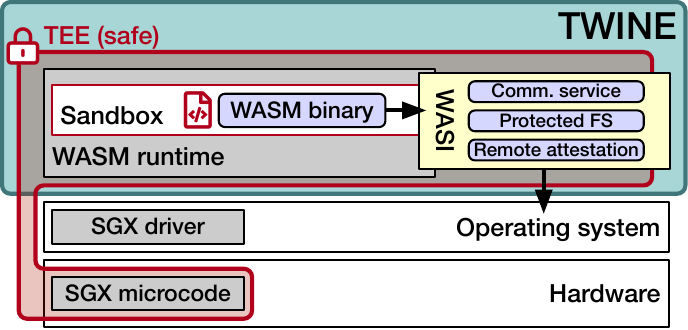}
    \caption{Overall \twine architecture.}
    \label{fig:arch}
\end{figure}

\subsection{Architectural Overview}
\twine comprises two main building blocks: a Wasm runtime and a WASI interface (see \Cref{fig:arch}). The Wasm runtime runs entirely inside the TEE, and WASI works as a bridge between trusted and untrusted environments, abstracting the machinery dedicated to communicate with the TEE facilities and the underlying OS.
The TEE-hosted Wasm runtime uses the WASI interface, which is always involved whenever OS services are accessed, allowing it to filter security-sensitive OCALLs.
Doing so serves as an intermediate control layer preceding the actual interaction with the OS, adhering to a capability-based security approach.
The runtime can limit what Wasm can do on a program-by-program basis, preventing Wasm code from leveraging the full access rights of the user owning the process.
For instance, WASI may restrict an application to only open a file system subtree, similar to the capabilities of \emph{chroot}.
The combination of the sandbox capabilities of SGX and WASI ends up in a two-way sandboxing system, partially inspired by MiniBox~\cite{minibox-atc14}, and somehow follows the same perspective of Ryoan.
The system, considered untrusted in the threat model of SGX, cannot compromise the integrity of the enclave code nor the confidentiality of the data stored in memory. Likewise, Wasm applications, considered untrusted from the data owner's standpoint, cannot interact directly with the OS unless WASI explicitly grants permission in the Wasm runtime.

\begin{figure}[!t]
    \centering
    \vspace*{-10px}
    \includegraphics[scale=0.6]{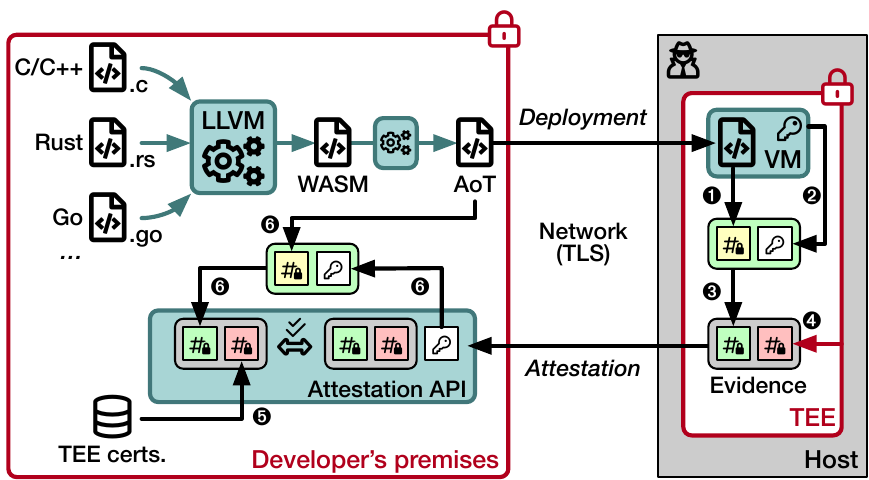}
    \caption{Overview of \twine's deployment and attestation workflow.}
    \label{fig:overview}
\end{figure}

\Cref{fig:overview} depicts the workflow of our proposal.
We compile source code from potentially multiple programming languages to Wasm bytecode and then perform AOT compilation to native code for enhanced performance.
We deploy these AOT-compiled applications via a secure communication channel to ensure code confidentiality and integrity.
Details on the attestation process for the hosted application and the runtime are elaborated in \S\ref{sec:ra}.

\paratitle{Compliance with R1} \twine facilitates the porting of legacy applications. In fact, it is language-independent. The programming language can be freely chosen, provided it can be compiled with LLVM or another compiler that supports Wasm and WASI as a compilation target. This lifts the restrictions imposed by SGX, typically forcing enclaved applications to be written in C/C++.
Furthermore, it is cross-platform hardware-compatible. Applications can be safely executed as long as the TEE is able to execute Wasm (supported by WASI), opening the door to other TEE technologies. Finally, it is system-agnostic as long as the OS can provide an equivalent of the API required by WASI. Since WASI mimics the system calls of POSIX systems, many Unix-like variants can implement it.

\paratitle{Compliance with R2, R4} We have appropriately instrumented the WASI interface and the runtime to control security-sensitive functionalities, which could be exploited to exfiltrate data. Depending on the selected configuration, low-level communication capabilities (socket-related OS calls) may not be provided directly to the Wasm binary. In such a case, communication over the network would only go through application-level protocols (\ie HTTP) to prevent covert channels via network system call interfaces. To this end, we embedded an HTTP library inside the Wasm module, which can be configured in terms of whitelisted targeted endpoints. 

\paratitle{Compliance with R3, R4} At the same time, file management support is provided and shielded against data leaks from code running in the Wasm module. The WASI interface has been extended so that every file operation leverages the \ipfs (IPFS), which automatically encrypts data coming out of the two-way sandbox (\eg via \texttt{fwrite}) and decrypts data flowing in the other direction (\eg via \texttt{fread}). 
Furthermore, the WASI sandbox enforces limitations on the Wasm applications by restricting them to predefined file system operations and access paths.

\paratitle{Compliance with R5} We also equipped \twine with an attestation service, which empowers data owners by enabling them to verify the authenticity and integrity of both the runtime environment and the WASI interfaces through which their data interacts. By implementing attestation mechanisms, data owners can confidently ensure that the runtime has not been tampered with and that the interfaces adhere to the specified security policies. This proactive approach to attestation fosters a foundation of trust between data owners and the runtime, assuring them that their sensitive information remains protected in the Wasm code and that their interactions with the runtime occur within a secure and verifiable environment, ultimately enhancing the overall security posture of the system. We equipped the Wasm module with an attestation library, which allows the data owner to generate an attestation quote whose \textit{EnclaveHeldData} field embeds the runtime hash, the security policy hash, and the wasm module hash. 

By default, Intel SGX ensures the integrity of the enclave binary rather than its confidentiality.
While integrity is verified through a signature of the code, the code itself must remain in plaintext to be loaded into enclave memory. Extensions, such as Intel SGX protected code loader (PCL), do provide confidentiality guarantees for enclave binaries, albeit with some security considerations, including writable sections and segments. 
These considerations may result in editable enclave code and read-only data at enclave runtime~\cite{IntelCorporation2020SGXDevRef}.
Conversely, \twine can also offer code confidentiality for enclave binaries without the limitations associated with PCL.
Wasm code can be either downloaded into the enclave following attestation or retrieved from a sealed blob and subsequently loaded by \twine. Upon decryption of the Wasm code, it is mapped into a secure memory area of SGX known as \emph{reserved memory}.
This memory region allows the runtime to load arbitrary executable code, and due to the inherent robustness of Wasm, such code remains immutable from the perspective of the Wasm application.

\section{Implementation Details}\label{sec:runti}

\subsection{Wasm Runtime and WASI}

As presented in \S\ref{sec:backg}, we considered many Wasm runtimes as candidates for implementing \sys.
We have chosen WAMR for its small size, few dependencies, and its ability to be linked to binary code (albeit generated ahead of time, that is, no JIT).
A small TCB reduces the attack surface of the runtime.
Further, AOT-compiled Wasm applications achieve near-native execution speed, as shown in \S\ref{sec:evalu}.
As such, we forked WAMR and extended its WASI interface, as explained below, in such a way that we can abstract the enclave constraints while implementing systems calls.

WASI is the interface through which Wasm applications communicate with the outside world, similar to POSIX's capabilities for regular native programs.
The development of TEE-enabled applications requires to deal with crossing the boundary between trusted and untrusted environments, materialised with \texttt{ECALL}s and \texttt{OCALL}s in the case of Intel SGX TEEs.
We believe that leveraging WASI as the communication layer meets the purpose of Wasm, where the implementation is abstracted away for the application itself.
As a result, the applications compiled in Wasm with WASI do not require any change to be executed inside Intel SGX or other TEE technologies.
For example, \textsc{WaTZ} showcased how the same Wasm applications can be hosted in TrustZone.

By the time \twine was developed, WAMR already included a WASI implementation that relies heavily on POSIX calls.
POSIX is not available inside SGX enclaves, so the WASI layer written by the authors of WAMR needs to cross the trusted boundary of the enclave frequently and straightforwardly routes most of the WASI functions to their POSIX equivalent using \texttt{OCALL}s.
While this approach enables running any Wasm applications that comply with WASI inside an enclave, it does not bring additional security benefits regarding the data that transits through POSIX, as there is no encryption.

We designed \twine to implement a more optimised WASI interface for WAMR, better tailored to SGX enclaves, which adopts a different approach than plain forwarding WASI calls outside the enclave.
The rationale for this choice is as follows.
First, performance: most WASI calls would simply be translated to (costly) \texttt{OCALL}s.
Second, we wanted to leverage trusted implementations when available, for instance, \ipfs (IPFS), described below (\S\ref{sec:ipfs}).
Therefore, we refactored WAMR's WASI implementation to keep its sandboxing enforcement, splitting the remaining into two distinct layers: \emph{(i)} one for specific implementations when available and \emph{(ii)} and another for generic calls.
Generic calls are handled by calling the POSIX library outside the enclave while providing additional security measures and validity checks.
Such calls are only wired when no trusted compatible implementation exists.
For instance, time retrieval is not supported by Intel SGX.
Hence, \sys's WASI layer fetches monotonic time while ensuring that the returned values are always greater than the previous ones.
If, for a given function, a corresponding trusted implementation exists (as it is the case for the ones in the official Intel SGX SDK), we use it to handle its related WASI call.
Often, a trusted implementation calls outside the enclave while at the same time providing additional security guarantees.
One notable example is the protected file system (see \S\ref{sec:ipfs}).
Finally, \sys can disable untrusted POSIX implementations inside the enclave (via a compilation flag). This is useful when one requires a strict and restricted environment or assesses how the applications rely on external resources.
In particular, the WASI interface may expose states from the TEE to the outside by leaking sensitive metadata in host calls, \eg usage patterns and arguments, despite the returned values being checked once retrieved in the enclave.

In its current implementation, \twine requires exposing a single \texttt{ECALL} to supply the Wasm application as an argument.
This function starts the Wasm runtime and executes the start routine of the Wasm application, as defined by WASI ABI specifications~\cite{wasiabi}.
\twine is versatile and can be adapted to only receive the Wasm applications from trusted endpoints supplied by the applications providers, as shown in \Cref{fig:overview}.
The endpoint may either be hard-coded into the enclave code and, therefore, part of the SGX measurement mechanism that prevents binary tampering, or provided in a manifest file with the enclave.
The endpoint can verify that the code running in the enclave (\ie the runtime) is trusted using SGX's remote attestation.
We propose a remote attestation API for Wasm applications in \Cref{sec:ra}.
As a result, \twine is a secure deployment and execution framework for running applications on untrusted devices and environments.
Despite OS dependency for network communication, \twine provides cryptographic techniques to create TLS and HTTPS channels that cannot be eavesdropped on.
For that purpose, we compiled and integrated a Wasm cryptographic library in the runtime (see \S\ref{sec:network}).

\subsection{Memory allocation}
Memory management greatly impacts the performance of the code executed in enclaves (see \S\ref{sec:polybench} and \S\ref{sec:sqlite:breakdown}).
WAMR provides three modes to manage the memory for Wasm applications:
\emph{(1)}~the default memory allocator of the system,
\emph{(2)}~a custom memory allocator, and
\emph{(3)}~a buffer of memory.
We found that using the SGX memory allocator to enlarge the linear memory of the Wasm runtime performed poorly, leading to a time complexity above linear.
Consequently, \sys preallocates a buffer of a fixed size to operate.
This approach is generally not problematic, as it requires specifying a fixed heap size at compile-time for SGX enclaves.

We note that the newer iterations of Intel SGX include a memory allocation scheme called enclave dynamic memory management (EDMM)~\cite{intel2016edmm}, enabling more memory allocation than the size specified at build time.
In such cases, \twine can leverage this new capability to extend the preallocated buffer dynamically and seamlessly.

\subsection{Communication Support}\label{sec:network}
Bringing network capabilities inside enclaves is essential to support communication with external endpoints, such as trusted peers or other enclaves.
Therefore, we implemented the required calls to deal with network sockets in our WASI layer, relying on the network stack of the untrusted OS.
We perform our cryptography inside the enclave to secure the communication channels and prevent attackers from eavesdropping.
For that purpose, we use WolfSSL~\cite{wolfssl}, an open-source popular cryptographic library.
WolfSSL supports mainstream ciphers and the TLS protocol, which can be used to set up trusted communication channels.
Using a renowned cryptographic library compiled in Wasm has many advantages: 
\emph{(1)}~the library is platform-independent and reusable in other TEEs (\eg TrustZone with \textsc{WaTZ}~\cite{menetrey2022watz}), 
\emph{(2)}~the library can be statically linked to any application when compiled into the Wasm format, and 
\emph{(3)}~the library can also leverage the multi-module feature of WAMR, the runtime which \sys is based on, which enables Wasm applications to load dependencies dynamically (\ie at runtime), which eliminates the burden of static linking, and abstracts a specific implementation of the library.
A WASI proposal already exists to bring cryptography to Wasm applications by the runtime. However, we considered its status too preliminary to be considered as a building block.

Furthermore, we adapted Mongoose~\cite{mongoose}, a lightweight and embeddable Web server library, to enable compilation into Wasm and facilitate the hosting of Web applications within the TEE.
In conjunction with WolfSSL, our adaptation of Mongoose enables clients to establish secure communication channels featuring HTTPS termination secured by Intel SGX.
Consequently, enclaved applications can expose high-level APIs, such as REST, while ensuring the confidentiality of data and the integrity of executing code.
The literature has examined various approaches to providing attestation in conjunction with high-level APIs supported by TLS.
Intel proposed an X.509 certificate extension, introducing specific object identifiers (\ie fields) to bind TLS and attestation~\cite{knauth2018integrating}.
Other researchers have extended the TLS handshake to incorporate additional attestation information~\cite{10.1145/1456455.1456462}.
\sys can employ one of these solutions for TEE attestation.

The WASI calls related to the sockets are implemented using \texttt{OCALL}s, forwarding them to the untrusted OS.
Analogous to the WASI implementation for Linux, our approach supports the sandboxing of networking, allowing the enclave launcher to supply IP ranges to which Wasm applications can connect.
We minimised the number of \texttt{OCALL}s by embedding computations that do not require untrusted OS system calls, \eg, text to binary IP address conversions.
WolfSSL has been slightly adapted to be compiled in Wasm.
Our work builds upon the compilation target of WolfSSL for Intel SGX, with missing dependencies addressed using WASI calls and WASI-SDK header files.
Due to the constraints of the Wasm virtual machine, which prohibits embedding assembly instructions in the bytecode, we could not use the hardware acceleration offered by modern CPUs for specific ciphers (\eg AES).
Limitations can be mitigated by offloading certain cryptographic operations to the runtime.
Mongoose exclusively supports OpenSSL and Mbed TLS libraries for providing cryptographic primitives.
As such, we integrated WolfSSL as a TLS provider within Mongoose, enabling it to host or query HTTPS-enabled websites.
We plan to contribute these changes to WolfSSL's and Mongoose's repositories.

\subsection{File Management Support}
\label{sec:ipfs}

As a showcase of the abstraction power offered by WASI, we implemented the subset of the WASI calls related to file system operations by using the \ipfs (IPFS)~\cite{ipfs}.
Being shipped with the Intel SGX SDK, it mimics the POSIX functions for file input/output.
The architecture of IPFS is split in two: \emph{(1)} the trusted library, running in the enclave that offers a POSIX-like API for file management, and \emph{(2)} the untrusted library, an adapter layer to interact with the POSIX functions outside of the enclave, that actually read and write on the file system.
Upon a \texttt{write}, content is encrypted seamlessly by the trusted library before being written on the media storage from the untrusted library.
Conversely, content is verified for integrity by the enclave during \texttt{read}.

IPFS uses AES-GCM for authenticated encryption, leveraging the CPU's hardware acceleration. An encrypted file is structured as a Merkle tree with nodes of a fixed size of \qty{4}{\kibi\byte}. 
Each node contains the encryption key and tag for its children nodes.
Thus, IPFS iteratively decrypts parts of the tree as the program in the enclave requests data~\cite{IPFSexplained}.
This mechanism ensures the confidentiality and integrity of the data stored at rest on the untrusted file system.
While the enclave is running, the confidentiality and the integrity of the data are also guaranteed by SGX's memory shielding.

IPFS has several limitations, which are deemed beyond the scope of Intel's threat model.
Since the files are saved in the regular file system, there is no protection against malicious file deletion and swapping.
Consequently, IPFS lacks protection against:
\emph{(1)}~rollback attacks: IPFS cannot detect whether the latest version of the file is opened or has been swapped by an older version, and
\emph{(2)}~side-channel attacks: IPFS leaks file usage patterns and various metadata such as the file size (up to \qty{4}{\kibi\byte} granularity), access time and file name.
We note how Obliviate~\cite{ahmad2018obliviate}, a file system for SGX, partially mitigates such attacks.
Although Obliviate can be adapted for use with \twine, addressing side-channel attacks falls beyond our threat model.

The WASI API includes several calls that do not have direct counterparts in the IPFS, due to slight variations from the ISO C standard.
For instance, \texttt{fseek} allows the cursor to move past the end of a file, which is not permitted in IPFS.
To address this, our WASI implementation extends the file with \texttt{null} bytes, requiring extra IPFS calls.
Also, IPFS lacks support for vectored read and write operations.
Since WASI exclusively handles file I/O with vectored operations, we resolved to implement those with an iteration.

IPFS provides convenient support to automatically create keys for encrypting files, derived from the enclave signature and the processor's (secret) keys.
While automatic key generation seems straightforward, a key generated by a specific enclave in a given processor cannot be regenerated elsewhere.
IPFS circumvents this limitation with a non-standard file open function, where the caller passes the key as a parameter.
Our prototype relies on automatic generation, and we leave it as future work to extend the SGX-enabled WASI layer to support custom encryption keys.

In conclusion, files persisted by \twine cannot be read outside the enclaves and are transparently decrypted and integrity-checked while handled by Wasm applications.

\subsection{Attestation}\label{sec:ra}
RA is a cornerstone feature of TEEs, as it ensures the authenticity of the executing code, including Wasm applications in the context of \sys.
We worked with the open-source community of WAMR~\cite{Menetrey2022rareq,Zeuson2022rapr} to define additional functions in the runtime to interact with the attestation features of Intel SGX.
Consequently, Wasm applications within our system can interface with Intel SGX to generate quotes during attestation.
A quote refers to a signed data structure that contains the enclave's measurement, identity, and additional metadata, certifying the enclave's trustworthiness, integrity, and authenticity to remote parties or Wasm applications executing in other SGX enclaves.
The integration of attestation within the runtime provides robust security guarantees for remote peers, typically facilitating the establishment of secure communication channels to transfer confidential data in remotely executed Wasm applications.

The RA feature of the Wasm runtime is implemented using \texttt{librats}, a low-level library facilitating attestation for multiple TEEs~\cite{InclavareContainers2023librats}.
Although the current implementation supports Intel SGX data centre attestation primitives (DCAP), the runtime can be extended to incorporate additional TEEs.
\Cref{fig:overview} illustrates the attestation workflow.
A hash is computed upon loading the Wasm application bytecode (or the AOT-compiled assembly code) within the enclave (\DingBlack{1}).
When collecting a quote, the runtime retrieves this precomputed hash and derives a secondary hash, incorporating optional user data provided by the Wasm application, such as a public key to establish a trusted communication channel (\DingBlack{2}--\DingBlack{3}).
The runtime then forwards the final value to the Intel SGX RA mechanism, which generates a quote (\DingBlack{4}).
This quote serves as evidence for a relying party, confirming the enclave's genuineness (\DingBlack{5}) and the trustworthiness of the executing Wasm application (\DingBlack{6}).

\section{\hspace{-5pt}\resizebox{240pt}{!}{Use Case: Credit Scoring in Crypto Finance}}\label{sec:usecases}
The \twine runtime has been used in a data-intensive, large-scale commercial application to enhance the trustworthiness of a distributed credit scoring oracle.
\emph{Credit scoring} has been historically used by financial institutions to estimate the risk of lending money to an individual. 
It determines the ability of an entity to repay debts based on a number of quantitative and qualitative metrics. 
High credit scores lead to higher chances of obtaining a loan with low interest rates. 
Conversely, entities with a lower credit score must pay higher interest rates on their loans. 
The \xmi company provides a privacy-preserving scoring solution for credit in crypto-currency finance. 
With over \$$100$B of crypto-collateral being used to generate over \$$1.25$B of interest on a quarterly basis, credit is one of the most rapidly growing sectors of the emerging crypto-currency finance ecosystem. 
\xm allows borrowers to supply lenders with real-time portfolio risk metrics, while preserving the privacy of trades, positions, and other sensitive information. 
Borrowers benefit from improved lending terms, as they can display their risk in real-time and assure lenders they are trading responsibly. 
Lenders benefit from increased visibility and real-time information. 
\xm calculates various dynamic risk metrics and aggregates across client portfolios, including total assets, liabilities and maximum loss simulations.
The latter is based on the standard portfolio analysis of risk (SPAN) system, wherein the worst possible loss of a
client's portfolio is estimated from a simulation over price and volatility shock scenarios.
The fundamental requirement of the \xm solution is that users' confidential data remains private and is computed correctly, \ie metrics must reflect the actual status of the credit.
To this end, \xm uses TEE technologies and cryptographic proofs to ensure users' sensitive data privacy and guarantee risk analysis. 
Intel SGX is the enabling TEE technology adopted to protect and attest sensitive processing. 
Within the threat model supported by Intel SGX, only computations authorised by the user are allowed in the attested enclave, and no party can see granular private data or perform any knowledge extraction. 

\begin{figure}[!t]
    \centering
    \includegraphics[width=\linewidth]{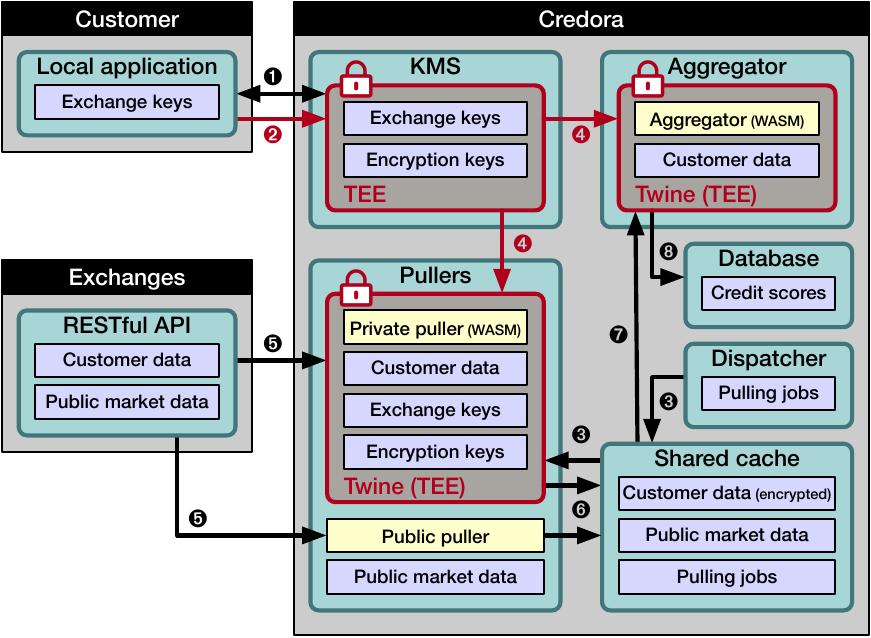}
    \caption{Architectural overview and workflow of \xm, highlighting attested channels and \twine enclaves in red.}
    \label{fig:xmargin}
\end{figure}

Figure~\ref{fig:xmargin} shows the architecture of the credit scoring system of \xm. 
It is composed of a set of distributed and loosely-coupled microservices communicating via a distributed in-memory data store.
The typical execution flow is as follows.
Initially, before interacting with the backend, the client challenges the KMS for its attestation  (\DingBlack{1}) to set up a secure channel.
Once completed, it can provide secrets (\ie \emph{exchange keys}) to the \xm TEE-secured KMS (\DingRed{2}). 
The \emph{dispatcher} defines pulling jobs, \ie a request to be executed for a particular exchange using its API endpoints that returns information on clients' portfolios. 
These jobs are transferred through a shared cache to the pullers (\DingBlack{3}).
Based on those, the \emph{private puller} uses the \emph{exchange keys} distributed by the KMS (\DingRed{4}) to get clients' data (\eg information on open trading positions) from crypto-currency exchanges venues (\DingBlack{5}) such as \emph{Binance}, \emph{Deribit}, \emph{Coinbase} and \emph{Kraken}.
A trading position denotes an individual's or entity's ownership stake in a particular financial asset, which represents their investment and potential for profit or loss.
The obtained data is encrypted and pushed into the shared cache (\DingBlack{6}).
Similarly, the \emph{public puller} gathers public market data (\DingBlack{5}), which is also stored in the shared cache (\DingBlack{6}).
The \emph{aggregator} reads clients' private data from the shared cache, markets public data obtained by the \emph{public puller} (\DingBlack{7}), and computes the risk metrics and credit scores (\DingBlack{8}).

In this use-case, the confidential data is given by: \emph{(1)}~\emph{exchange keys} used to obtain clients' wallet data from \emph{exchange} venues, and \emph{(2)}~the client trading positions received from \emph{exchanges}.
This architecture guarantees that confidential data never leaves the secure enclave unencrypted. The \emph{exchange keys} are received directly from clients' browsers over an attested TLS TEE-terminated secure channel. This is possible through the \emph{librats} library~\cite{InclavareContainers2023librats}, which can be compiled to Wasm with \emph{emscripten} and executed inside the browser.
These \emph{exchange keys} are persistently stored using IPFS that encrypts the secret information in the enclave using the SGX sealing key. 
The process of signing requests for \emph{exchanges} using the \emph{exchange keys} is executed inside the enclave. 
RESTful requests for \emph{exchanges} are then made over an HTTPS TEE-terminated connection using WolfSSL and Mongoose compiled in Wasm, guaranteeing that the confidential financial data is directly received and aggregated in the enclave. 

Under the described scenario, \xm's clients only need to trust Intel (\ie SGX threat model) and \xm itself, which does not disclose the implementation of their software.
Hence, the cloud provider hosting the application may be untrusted, thanks to the isolation offered by Intel SGX.
Additionally, we aim to further reduce the scope of the threat model by removing \xm, leaving Intel as the only trusted entity.
Toward this goal, one must prove to clients that no data has ever leaked from the secure enclaves, nor that the enclaves are curious and retrieve unwanted information from the cloud provider's system.
As such, \sys's trust model plays an essential role: its two-way sandbox (as explained in \S\ref{sec:runti}) offers two strong guarantees for the cloud provider and the clients.
First, the \xm's enclaves, which are based on \sys, are proven authentic using RA, which guarantees to \xm that the enclaves have not been tampered with and can supply Wasm applications and confidential information for further computations, preventing one from eavesdropping.
Second, \xm's customers can review the open-source implementation of \xm's enclaves (\ie based on \sys), ensuring that the runtime properly sandboxes the Wasm applications (whose source code is proprietary) deployed by \xm later on, which prevents the \xm's Wasm applications from accessing the host system resources.

\section{Evaluation}\label{sec:evalu}

We present here our extensive evaluation of the runtime \twine.
We intend to answer the following questions:
\begin{itemize}
    \item What are the performance overheads of using \sys on Client and Scalable SGX, compared to native applications and \acctee, a state-of-the-art solution?
    \item What are the performance overheads for using cryptographic operations and setting up TLS-terminated connections within the enclaves?
    \item Can a database engine be compiled into Wasm and executed in a TEE while preserving acceptable performance? 
    \item How do the database input and output operations behave when the EPC size limit is reached?
    \item What are the primitives that generate most of the performance overheads while executing database queries? Can we improve them?
    \item How does \sys perform when used in a data-intensive and real-world solution?
\end{itemize}

We answer these questions by using a general-purpose compute-bound evaluation with \polybench (\S\ref{sec:polybench}), 
encrypting, hashing and securing communications using cryptographic primitives and TLS with WolfSSL (\S\ref{sec:networkstack}), 
evaluating a general-purpose embeddable database using SQLite (\S\ref{sec:speedtest1}), 
stressing the database engine using custom micro-benchmarks that perform read and write operations (\S\ref{sec:sqlite:breakdown}), 
analysing various cost factors bound to Wasm and SGX (\S\ref{sec:cost-factors}), 
profiling the time breakdown of the database components, the Wasm runtime and the SDK of SGX (\S\ref{sec:profiling}), 
and finally assessing the end-to-end performance of a Wasm application in the fintech company \xm (\S\ref{sec:xm}).

\subsection{Experimental setup}\label{sec:envsetup}
We use a Supermicro 5019S-M2 with Intel Xeon E3-1275 v6 (\qty{3.8}{\giga\hertz}, EPC \qty{128}{\mebi\byte}, usable \qty{93}{\mebi\byte}) for Client SGX tests, and a Supermicro SYS-520P-WTR with an Intel Xeon Gold 6326 (\qty{2.9}{\giga\hertz}, EPC \qty{8}{\gibi\byte}) for Scalable SGX.
The benchmarks are executed using Client SGX, except where noted.
Systems run Ubuntu 18.04.6 with kernel 4.15.0-202, SGX driver v2.11, and SGX SDK v2.17.100.3.
\sys is fully merged into the WAMR's official repository.
As such, we use the WAMR build for running the benchmarks unless stated otherwise.

Time is measured using the monotonic clock of POSIX in all the benchmarks and averaged using the median.
If measured from within the enclave, the time to leave and reenter the enclave is included.
In our setup, the enclave round trip accounts for approximately \qty{4}{\milli\second}.
We used Docker to build the benchmarks while their execution is on bare metal to avoid potential isolation overheads.
The native benchmarks are compiled using Clang 10 with optimisation set to \texttt{--O3}.
The Wasm benchmarks are compiled using Clang into Wasm format, then AoT-compiled into native format using the compiler provided by WAMR (\ie \texttt{wamrc}) using \texttt{--O3} and size level 1 to run into SGX enclaves (\texttt{--sgx}).
We used GCC v7.5 for two tasks: \emph{(1)} compile the applications executing the benchmarks, \ie the WAMR runtime and the SGX enclaves, also with \texttt{--O3}, and \emph{(2)} compile IPFS with \texttt{--O2}, as in the SGX SDK.
SGX-LKL (v0.2.0), LKL (v5.4.62) and \acctee have been used as an empirical baseline for running the experiments natively in SGX enclaves.
They have been downloaded from the official Debian repository and GitHub. 
Finally, our implementation and instructions to reproduce our experiments are open-source~\cite{twine-github}.

\subsection{Micro-benchmarks: \polybench}\label{sec:polybench}

\begin{figure*}[!t]
    \centering
    \includegraphics{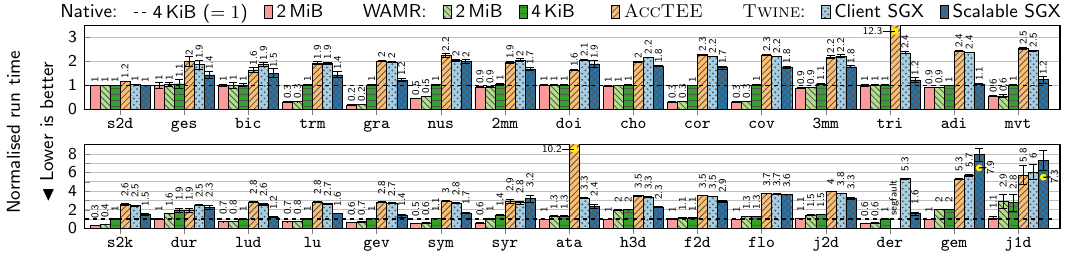}
    \vspace{-5pt}
    \caption{Performance of \polybench benchmarks, normalised to native speed, with memory page sizes of \qty{4}{\kibi\byte} and \qty{2}{\mebi\byte}.}\label{fig:polybench}
\end{figure*}

\polybench\cite{pouchet.11.polybench} is a CPU-bound benchmark suite commonly used to validate compiler optimisations and compare the performance of Wasm execution environments~\cite{DBLP:conf/usenix/JangdaPBG19,DBLP:conf/middleware/GoltzscheNKK19}.
We leveraged \polybench due to the practicality of deploying it in SGX enclaves.
We present results for 30 \polybench (v4.2.1) tests: native (x86-64 binaries), AOT-compiled Wasm using WASI-SDK for WAMR and \twine, and JIT-compiled using V8 for \acctee.
We also benchmark \twine with Client and Scalable SGX, highlighting performance overheads due to the EPC limit.
For accurate precision when compared to \acctee, a state-of-the-art runtime, we did not instrument the Wasm bytecode.
Furthermore, we benchmarked native and WAMR with two sizes of memory pages: \qty{4}{\kibi\byte} and \qty{2}{\mebi\byte}.
Intel SGX-enabled solutions omit large memory pages due to the absence of support in SGX.
\Cref{fig:polybench} shows the results normalised against the native run time using memory pages of \qty{4}{\kibi\byte}.

We can split the \polybench test results into four groups based on the proportion between the execution modes (native, WAMR and \sys) and the memory page sizes (\qty{4}{\kibi\byte} and \qty{2}{\mebi\byte}): 
\emph{(1)}~similar execution time (\eg \texttt{s2d}); 
\emph{(2)}~WAMR, \acctee and \sys (\ie Wasm) are slower than native (\eg \texttt{h3d}, \texttt{gem} and \texttt{j1d}); 
\emph{(3)}~\acctee and \twine (\ie Wasm coupled to SGX) are slower than WAMR and native (\eg \texttt{ges}, \texttt{bic} and \texttt{doi}); 
\emph{(4)}~native and Wasm using large memory pages are faster than native and Wasm using regular memory pages (\eg \texttt{trm}, \texttt{gra} and \texttt{cor}).

While Wasm may be up to speed with native performance, Wasm is usually slower than native due to several reasons: increased register pressure, more branch statements, increased code size, high reliance on the stack, \etc.
We also looked at the impact of memory on performance, given the additional cost for SGX enclaves~\cite{199364}.
Beginning with \qty{160}{\mebi\byte}, the minimum to initiate all \polybench tests, we incrementally reduced the Wasm runtime memory allocation until memory allocation failed.
We found that the slowdown in the \texttt{der} test is due to reaching the EPC size limit.
\acctee notably throws a segmentation fault for this test unless the array dimensions are reduced (as in the original paper).
Similarly, \texttt{lu} and \texttt{lud} require at least \qty{80}{\mebi\byte} of memory.
In our settings, \acctee showed spikes in performance for the \texttt{tri} and \texttt{ata} tests.
Despite higher normalised run times compared to \twine, the absolute run times for these tests stay under one second.

We further examine the overheads caused by the page evictions from the EPC.
For that purpose, we run Polybench/C on \twine with Scalable SGX, benefiting from a larger EPC.
Memory-intensive tests, such as \texttt{der}, \texttt{lu}, and \texttt{lud}, reveal fewer overheads compared to Client SGX.
Besides, the majority of tests show reduced overheads, even if they did not use the total EPC capacity.
A possible reason for this performance difference is Scalable SGX's transition to Intel's Total Memory Encryption, supported by AES-XTS~\cite{intel2021tme}, over the traditional Merkle tree that maintained the integrity of enclaves.
While this choice downgrades SGX's threat model by relaxing defences against hardware integrity and anti-replay attacks, this trade-off offers larger EPCs and generally more efficient enclave execution~\cite{johnson2021supporting}.

We also analysed the usage of large memory pages for two reasons.
First, enabling large pages for a process that has yet to be tailored for using them usually requires turning on the OS-wide feature \emph{Transparent Huge Pages}, impacting all the other processes running on the same system.
On the other hand, \sys abstracts the memory allocator by acting as a middleware between the OS and the Wasm application and may allocate large pages for a given Wasm program granularly.
Second, \polybench highlights that Wasm memory-intensive workloads with large pages enabled may be faster than their native counterparts when not configured for using this feature, offering an effortless performance boost.
This speedup is due to the reduced pressure on the processor translation lookaside buffer (TLB) cache, which translates virtual to physical memory addresses.

\begin{figure*}[!htb]
    \centering
    \includegraphics{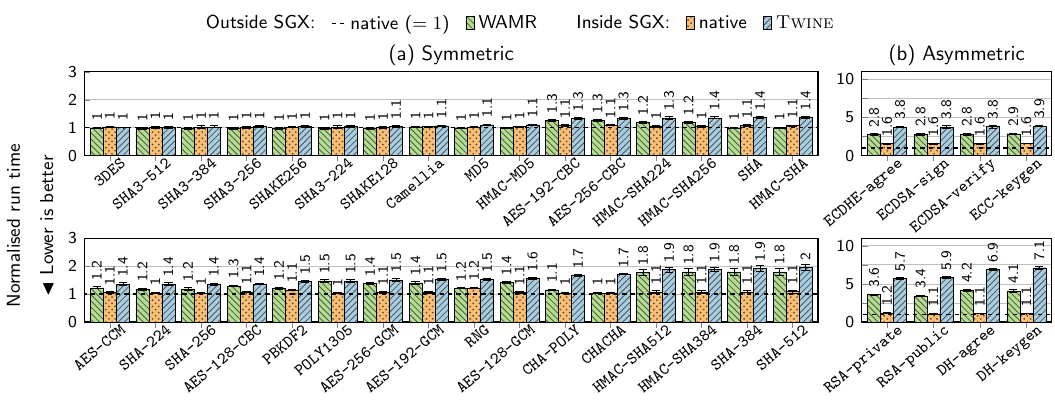}
    \caption{Performance of WolfSSL benchmarks targeting cryptographic algorithms, normalised to the native speed.}
    \label{fig:crypto}
    \vspace{-12pt}
\end{figure*}
\begin{figure}[!htb]
    \centering
    \includegraphics{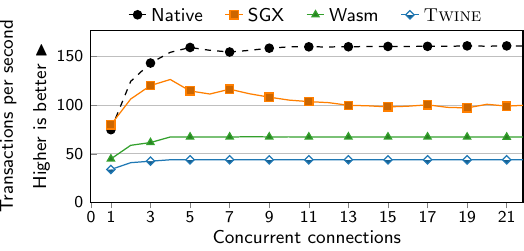}
    \caption{Performance of WolfSSL benchmarks for TLS sessions.}
    \label{fig:tls}
\end{figure}

\subsection{Micro-benchmarks: network stack}\label{sec:networkstack}

We assess the network stack of \sys using two micro-benchmarks bundled with WolfSSL: 
\emph{(1)} a performance comparison of cryptographic primitives using many ciphers and hashing algorithms, and 
\emph{(2)} a performance evaluation of TLS sessions, which have one side of the connection terminated within the enclave.
In both cases, we compare the relative execution speed of native and Wasm (inside SGX) and Wasm (outside SGX) against native (outside SGX).
For a fair comparison between native and Wasm, we have chosen to disable the hardware acceleration support of WolfSSL, \ie the offloading of some ciphers using CPU capabilities.

Figure \ref{fig:crypto}a depicts the execution speed of symmetric algorithms and hashing functions.
In contrast, Figure \ref{fig:crypto}b illustrates the speed for asymmetric ciphers and key generation operations.
Among these, symmetric operations are the most efficient, with an average slowdown for Wasm of \cryptoSymRatioWasmToNative$\times$ outside of SGX, and \cryptoSymRatioWasmSgxToNative$\times$ inside of the TEE, compared to native speed execution.
Hashing functions follow, with slowdowns of \cryptoHashRatioWasmToNative$\times$ and \cryptoHashRatioWasmSgxToNative$\times$ for Wasm outside and inside SGX, respectively.
Finally, the asymmetric ciphers have the highest slowdown, with \cryptoAsymRatioWasmToNative$\times$ and \cryptoAsymRatioWasmSgxToNative$\times$ for the same settings.
The TLS protocol uses asymmetric ciphers to establish sessions, authenticate remote endpoints, and exchange session keys.
Once the session is active, it relies on more efficient symmetric and hashing algorithms, mitigating concerns over the slower speed of asymmetric ciphers.

In Figure \ref{fig:tls}, we stressed an application using WolfSSL's TLS 1.3 protocol by evaluating the number of TLS transactions per second over a range of concurrent connections.
The setup involves a client and a server hosted on different machines connected through a switch.
The client is a native executable running on Linux, while the server is of four types and evaluated separately: native in Linux and SGX, Wasm in Linux and \sys in SGX.
A TLS transaction is comprised of \emph{(1)} the TLS handshake, \emph{(2)} the server reading \qty{16}{\kibi\byte}, \emph{(3)} the server sending \qty{16}{\kibi\byte}, and \emph{(4)} the closure of the session.
We measure the time the server takes to handle \num{512} connections while varying the number of concurrent connections handled by a single core.
Besides, we used the ciphersuite \texttt{TLS13-AES128-GCM-SHA256} as cryptographic primitives.
Finally, we considered \texttt{AES128-CCM} and \texttt{CHACHA20-POLY1305} as AEAD ciphers, but we did not notice significantly different results because the performance of these algorithms is similar, as reflected in Figure \ref{fig:crypto}.
\begin{figure*}[!t]
    \centering
    \includegraphics{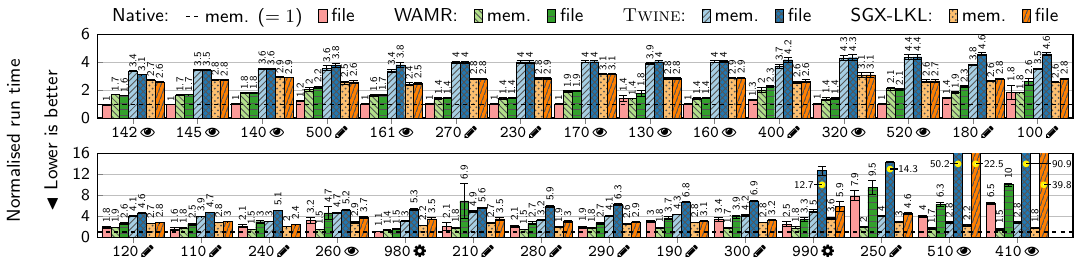}
    \caption{Relative run time of SQLite \texttt{Speedtest1} benchmarks. (\faEye\:=\:reading, \faPencil\:=\:updating, \faGear\:=\:housekeeping).} 
    \label{fig:speedtest}
    \vspace{-12pt}
\end{figure*}
The number of transactions per second for native execution outside of SGX and Wasm demonstrates a pronounced increase until it reaches a maximum value, after which it stabilises.
This maximum value signifies the saturation of the singular thread responsible for managing TLS sessions.
The native execution within the enclave exhibits a comparable pattern; however, it necessitates a more extended period to converge after attaining its peak value.
Although we did not conduct a comprehensive investigation on this particular behaviour, we believe the observed phenomenon results from how SGX-LKL (the library operating system employed for the execution of native applications in SGX) processes in-enclave packets via its dedicated TCP/IP stack.
Native execution outside of SGX converges to an average of \qty{\tlsAverageNative}{\tps}, which serves as the baseline measurement.
In contrast, native execution within the TEE exhibits an average of \qty{\tlsAverageSgx}{\tps} with a consequent slowdown factor of \tlsRatioSgx$\times$.
Furthermore, the average transaction rates per second for Wasm outside and inside the enclave are \qty{\tlsAverageWasm}{\tps} and \qty{\tlsAverageTwine}{\tps}, respectively, with corresponding slowdown factors of \tlsRatioWasm$\times$ and \tlsRatioTwine$\times$.
When comparing the in-enclave solutions, \sys's slowdown relative to native is \tlsRatioSgxVsTwine$\times$, but offers all the advantages of Wasm, such as portability and security.
Yet, Wasm services hosting TLS connections may leverage multithreading for performance enhancement.

We highlight that modifications were made to the official benchmark to use the \texttt{poll} system call instead of \texttt{epoll}, as the latter is unsupported in WASI.
The resulting software was then contributed to the WolfSSL open-source repository.

\subsection{Macro-benchmarks: SQLite}\label{sec:speedtest1}

SQLite~\cite{5231398} is a widely-used full-fledged embeddable database.
It is ideally suited for SGX, thanks to its portability and compact size.
For this reason, we thoroughly evaluated it as a showcase for performance-intensive operations and file system interactions.
SQLite requires many specific OS functions missing from the WASI specifications due to standardisation and portability concerns in Wasm.
Therefore, we leveraged SQLite's virtual file system (VFS) and implemented a minimal file system interface compatible with WASI to make SQLite process and persist data, reducing the POSIX functions to be supported.
We used one of the official templates (\texttt{test\_demovfs}) to override the OS interface of SQLite since it relies on a few POSIX functions covered by the WASI specifications.
SQLite uses a \num{2048}-page cache of \qty{4}{\kibi\byte} each (for a cache size of \qty{8}{\mebi\byte}) with the default (normal) synchronous mode and the default (delete) journal mode.
Besides, we use an alternate memory allocator  (\texttt{SQLITE\_ENABLE\_MEMSYS3}) to provide a large chunk of preallocated memory used for the database instance.

Memory allocation in SGX enclaves can be costly, consuming up to 45\% of CPU time in some tests.
Preallocating memory can, therefore, significantly improve performance when the database size is predictable.
We executed SQLite v3.32.3-amalgamation (\ie a single-file version of the entire SQLite program).
First, we used SQLite's own performance test program, \texttt{Speedtest1}~\cite{speedtest1}, running 29 out of the available 32 tests, covering a large spectrum of scenarios (we excluded three experiments because of issues with SQLite VFS).
Each \texttt{Speedtest1} experiment targets a single aspect of the database, \eg selection using multiple joints, the update of indexed records, \etc.
Tests are composed of an arbitrary number of SQL queries, potentially executed multiple times depending on the load to generate.
\Cref{fig:speedtest} shows our results, normalised against the native execution.
We include results for in-memory configurations as well as for a persisted database, where WASI is used.

While we provide additional details below, we observed across all tests that the WAMR's slowdown relative to native on average is \speedtestWamrMemToNativeRatio$\times$ for in-memory and \speedtestWamrFileToNativeRatio$\times$ for file-based databases.
\sys's slowdown relative to WAMR is \speedtestTwineMemToWamrRatio$\times$ for in-memory and \speedtestTwineFileToWamrRatio$\times$ for file-based databases.
Symbols (\faEye, \faPencil, \faGear) indicate read queries, data updates (\eg inserting, updating, and deleting records), and housekeeping tasks (\eg integrity checks, statistics collection).
The upper plot shows a consistent performance penalty trend by variant, with a slowdown of \speedtestTwineMemRatioUpperPlot$\times$ and \speedtestTwineFileRatioUpperPlot$\times$ for \sys in-memory and file-based databases, respectively.
These experiments primarily represent read queries from the benchmarks.
Experiments demonstrating identical performance for in-memory and persistent databases suggest actions on the page cache without file system interaction.
Utilising SGX with a persistent database incurs substantial overhead under specific conditions.
Notably, experiments 410 and 510 induce additional latency attributed to intensive I/O operations.
Specifically, each query in these tests repeatedly calls the libc function \texttt{fstat} to get the database file size, an operation further slowed by enclave \texttt{OCALL}s. 
We note that caching this information at the WASI layer could improve the performance of these recurring calls, provided that the file remains unmodified.
Compared to the 410 test on an in-memory database, \sys and SGX-LKL reveal slowdowns of up to \speedtestExpFourOneZeroTwineMemVsFileRatio$\times$ and \speedtestExpFourOneZeroSgxLklMemVsFileRatio$\times$, respectively.
Test 210 is I/O-intensive, as it modifies the database schema and, consequently, all records.
Moreover, test 250 is highly I/O-intensive within a persisted database, as it updates every record in a table, requiring the re-encryption of a significant portion of the database file.
Finally, 990 is a particular case of database housekeeping.
It gathers statistics about tables and indices, storing the collected information in internal database tables where the query optimiser can access and use the information to help make better query planning decisions.
The protracted execution time of \sys and SGX-LKL with a persistent database is attributed to the added complexity of I/O operations from the enclave.
In general, a consistent pattern is observed across variants, except in cases involving extensive file system interactions, which typically influence native execution outside of SGX as well.

\subsubsection{SQLite: overhead breakdown}\label{sec:sqlite:breakdown}

\begin{figure*}[!htb]
    \centering
    \includegraphics{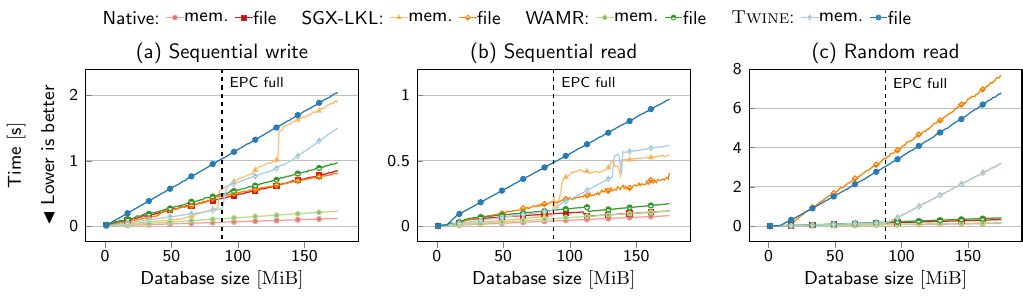}
    \caption{SQLite, microbenchmarks for sequential \texttt{write} (a), sequential \texttt{read} (b), random \texttt{read} (c). We compare four variants (native, SGX-LKL, WAMR, \sys) and two different modes (in memory and on file).}
    \label{fig:microbenchmarks}
    \vspace{-12pt}
\end{figure*}

In order to comprehensively examine the origins of observed performance penalties, we devised a test suite to assess prevalent database queries.
This suite encompasses various query types, including insertion, sequential, and random reading (measured independently due to their distinct complexities~\cite{199364}).
This test suite's design follows comparable benchmarks from previous literature~\cite{Sartakov2018Stanlite}.
The tests use a single table with an auto-incrementing primary key and a blob column.
For sequential insertions, the blob column is iteratively filled by an array of random data (\qty{1}{\kibi\byte}) using a pseudorandom number generator (PRNG, same as \texttt{Speedtest1}).
Next, records are selected in the order they have been inserted (\texttt{WHERE} clause).
Finally, we selected one random entry at a time.
The database is initialised with \qty{1}{\kilo\nothing} records (\ie \qty{1}{\mebi\byte} in total) and iteratively increases that amount by \qty{1}{\kilo\nothing} entries at the time, up to \qty{175}{\kilo\nothing} records (\ie \qty{175}{\mebi\byte}).
We evaluated four variants: a native version of SQLite running either outside or inside an enclave and an ahead-of-time Wasm version running with the same settings.
For each of them, we include results for in-memory and on-file databases.
The performance results for \sys (file-based) illustrate the enhanced version of IPFS, which reduces the latency of the read/write operations.
The details of the improvement of IPFS are covered in \S\ref{sec:profiling}.
\Cref{tbl:microbenchmarks} summaries the obtained results, where values on each line are normalised with the run time of the native variant.
The run time is the median of the queries' execution time, either from \qty{1}{\kilo\nothing} to \qty{175}{\kilo\nothing} records for native and WAMR, or split into two parts for SGX-LKL and \sys, going from \qty{1}{\kilo\nothing} to the EPC size limit (\textasciitilde \qty{88}{\mebi\byte}) and from that limit to \qty{175}{\kilo\nothing}.

\Cref{fig:microbenchmarks}a presents the results of the insertion of records.
While the variants executed outside the enclave demonstrate consistent performance, the EPC limitations impact the in-memory variants.
This outcome can be attributed to the costly nature of swapping operations associated with enclave memory pages, which involves the encryption of evicted pages~\cite{IntelCorporation2018SGXperf}.
Importantly, \sys demonstrates a significant performance improvement in the in-memory insertion above the EPC limit, with a gain of \insertLklVsTwineMemAepc$\times$.
This highlights the effectiveness of \sys in addressing the performance challenges associated with EPC constraints.
The operational cost associated with the persistent database in \sys escalates linearly because of file encryption.
Conversely, SGX-LKL employs a more optimal approach for inserting sequential elements and adheres to the performance trend exhibited by \sys's in-memory variant.

\newcommand{\twinebetter}{*}
\begin{table}[!t]
\vspace*{-1.5pt}
\small
\setlength{\tabcolsep}{2pt}
\caption{Comparison of the technologies in normalised run time.}
\rowcolors{1}{gray!0}{gray!10}
\begin{tabularx}{\columnwidth}{XS[table-column-width = 10mm]*4{S[table-column-width = 10mm, table-format = 2.1]}}
\toprule
\rowcolor{gray!25}
& &\multicolumn{2}{c}{SGX-LKL} &\multicolumn{2}{c}{\sys}\\[-3pt]
\rowcolor{gray!25}
&{\makecell[c]{\multirow{-2}*{WAMR}}} &{\raisebox{-1px}{\makecell[c]{\scriptsize{$<$EPC}}}} &{\raisebox{-1px}{\makecell[c]{\scriptsize{$\geq$EPC}}}} &{\raisebox{-1px}{\makecell[c]{\scriptsize{$<$EPC}}}} &{\raisebox{-1px}{\makecell[c]{\scriptsize{$\geq$EPC}}}}\\
\midrule
Insert mem.     &\insertWamrMem    &\insertLklMemBepc    &\insertLklMemAepc    &\insertTwineMemBepc                &\insertTwineMemAepc\twinebetter\\
Insert file     &\insertWamrFile   &\insertLklFileBepc   &\insertLklFileAepc   &\insertTwineFileBepc               &\insertTwineFileAepc\\
Seq. read mem.  &\readSeqWamrMem   &\readSeqLklMemBepc   &\readSeqLklMemAepc   &\readSeqTwineMemBepc               &\readSeqTwineMemAepc\\
Seq. read file  &\readSeqWamrFile  &\readSeqLklFileBepc  &\readSeqLklFileAepc  &\readSeqTwineFileBepc              &\readSeqTwineFileAepc\\
Rand. read mem. &\readRandWamrMem  &\readRandLklMemBepc  &\readRandLklMemAepc  &\readRandTwineMemBepc              &\readRandTwineMemAepc\\
Rand. read file &\readRandWamrFile &\readRandLklFileBepc &\readRandLklFileAepc &\readRandTwineFileBepc\twinebetter &\readRandTwineFileAepc\twinebetter\\
\bottomrule
\noalign{\vskip 2pt} 
\multicolumn{2}{l}{\cellcolor{gray!0}\scriptsize{(Native run time = 1)}} & \multicolumn{4}{r}{\cellcolor{gray!0}\scriptsize{\twinebetter\sys is faster than SGX-LKL.}}\\
\end{tabularx}
\label{tbl:microbenchmarks}
\end{table}

\Cref{fig:microbenchmarks}b illustrates the execution time required to read all records sequentially.
The variants executed outside the enclave exhibit linear costs, with a minor decline when the database accommodates \num{114000} records.
Although this study focused on \twine's performance, the slightly unexpected behaviour has yet to be further investigated.
Both \sys and SGX-LKL with an in-memory database display a pronounced increase beyond the EPC size limit, attributable to enclave paging.
In contrast, \sys with a file-based database demonstrates optimal performance while the database remains within the 8 MiB range (\ie the configured cache size for SQLite).
A similar increase is observed up to \qty{16}{\mebi\byte} (twice the cache size).
To confirm that this overhead is related to the cache, the cache size was increased to \qty{16}{\mebi\byte}, revealing that the sharp increase ceases at \qty{32}{\mebi\byte}.
We observe comparable trends by replacing the WASI layer with the one from WAMR, which does not employ encryption and relies on direct POSIX calls.
As a result, the primary source of the observed performance penalties can be attributed to SGX memory access operations.

\Cref{fig:microbenchmarks}c presents the execution time associated with random reading operations.
The costs for all variants increase linearly in proportion to the size of the database, with the exception of the SGX in-memory database variants, which are affected by EPC limitations.
Random reading operations trigger the enclave paging mechanism more frequently, as the spatial locality of the requested records no longer remains smaller than the size of a single memory page.
Notably, the in-file random reading scenario underscores the superior performance of \sys compared to SGX-LKL, exhibiting a \readRandLklVsTwineFileBepc$\times$ increase prior to the EPC limit and a \readRandLklVsTwineFileAepc$\times$ increase following the limit.

As a result, \sys exhibits somewhat lower performance compared to SGX-LKL, which can be attributed to the overhead associated with Wasm.
Nevertheless, \sys achieves competitive and even faster operations than SGX-LKL when it comes to random file access and in-memory insertion upon reaching the EPC threshold.
While \sys faces challenges in certain use cases, its overall performance remains robust.
A detailed examination of the role of SGX in this behaviour is provided in \Cref{sec:cost-factors}.

\begin{figure*}[!htb]
    \begin{minipage}[t]{\columnwidth}
        \centering
        \includegraphics{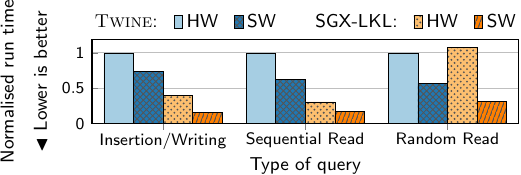}
        \caption{Comparison of normalised run time for SGX variants using in-file databases with SGX hardware and emulation.}
        \label{fig:execution-time}
    \end{minipage}
    \hfill
    \begin{minipage}[t]{\columnwidth}
        \centering
        \vspace*{-78px}
        \includegraphics{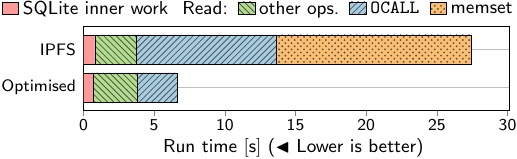}
        \caption{Run time breakdown prior to and following the optimisations of IPFS.}
        \label{fig:profiling}
    \end{minipage}
    \vspace{-12pt}
  \end{figure*}

\subsubsection{Cost factors assessment}\label{sec:cost-factors}

\def\labelcolumnlength{28mm}
\begin{table}[t]
\small
\setlength{\tabcolsep}{2pt}
\caption{Cost factors of the micro-benchmarks.}
\rowcolors{1}{gray!10}{gray!0}
\begin{tabularx}{\columnwidth}{p{\labelcolumnlength}RRRR}
\toprule
\rowcolor{gray!25}
\textit{(a)~Times {[ms]}} &Native                        &\textsc{sgx-lkl}                    &WAMR                        &\sys\\
\midrule
Compile runtime           & ---                          &\compilationSgxAverage              &\compilationWasmBinAverage  &\compilationWasmSgxBinAverage\\
Compile Wasm              & ---                          & ---                                &\compilationWasmAverage     &\compilationWasmAverage\\
Compile x86/AoT           &\compilationNativeAverage     &\compilationNativeAverage           &\optimizationWasmAotAverage &\optimizationWasmAotAverage\\
Gen. disk image       & ---                          &\optimizationSgxLklDiskImageAverage & ---                        & ---\\
Launch                    &\launchTimeNativeAverage      &\launchTimeSgxAverage               &\launchTimeWasmAverage      &\launchTimeWasmSgxAverage\\
\bottomrule
\rowcolor{gray!25}
\textit{(b)~Sizes {[KiB]}} &Native                       &\textsc{sgx-lkl}                    &WAMR                        &\sys\\
\midrule
Executable, disk           &\sizeNativeAverage           &\sizeSgxLklBinAverage               &\sizeWasmBinAverage         &\sizeWasmSgxBinAverage\\
Enclave, disk              & ---                         &\sizeSgxLklEnclaveAverage           & ---                        &\sizeWasmSgxEnclaveAverage\\
Wasm artifact, disk        & ---                         & ---                                &\sizeWasmAverage            &\sizeWasmAverage\\
AoT artifact, disk         & ---                         & ---                                &\sizeWasmAotAverage         &\sizeWasmAotAverage\\
Disk image                 & ---                         &\sizeSgxLklDiskImageAverage         & ---                        & --- \\
Executable, mem.         &\memoryNativeInMemoryAverage &\memorySgxInMemoryAverage           &\memoryWasmInMemoryAverage  &\memoryWasmSgxInMemoryAverage\\
Enclave, mem.            & ---                         &\memorySgxEnclaveSize               & ---                        &\memoryWasmSgxEnclaveSize\\
\bottomrule
\end{tabularx}
\rowcolors{1}{gray!10}{gray!0}
\label{tbl:cost-factors}
\end{table}

To evaluate the performance limitations and constraints associated with SQLite, we conducted a thorough analysis of the cost factors when using SGX and Wasm, either independently or conjointly.
We identified two distinct dimensions of cost implications:
\emph{(1)}~the time required to build and deploy an application, which transpires within the developers' domain, and 
\emph{(2)}~the time and storage space required to execute an application on an untrusted platform.
The analysis presented herein was executed on the preliminary version of \twine (before integration into the official repository).

\Cref{tbl:cost-factors}a summarises the time overheads we observed with the SQLite micro-benchmarks (\qty{175}{\kilo\nothing} records).
As different kinds of costs are involved depending on the variant, we do not indicate totals in the table.
The native one is composed of a single executable binary, while SGX-LKL requires the same executable binary and a disk image, which is an abstraction introduced to store the code and data securely.
The two variants that use Wasm require an executable (the runtime) and a Wasm artefact containing the SQLite benchmarks.
For both variants, we measured the time for AoT compilation as well.
For launching, we measured the time from the process creation to the start of the database initialisation.
The variants without SGX are naturally faster since they do not have to initialise the enclave.
The initialisation of \sys is \launchTimeTwineVsSgx$\times$ faster than SGX-LKL because the enclave is heavier than \sys's, and the benchmarks executable is encrypted on the disk image.

\Cref{tbl:cost-factors}b indicates the disk and resident memory footprints of the compiled components and additional prerequisite software.
The native variant is stored in a single executable binary file.
SGX-LKL has a heavier-sized executable and a much larger enclave binary.
The latter contains a generic program that is only loaded once and runs any other program stored in a disk image, such as our SQLite benchmarks.
A disk image is necessary for SGX-LKL, which it maps into RAM.
We generated an ext4-formatted file system, whose size is fixed at build time to be big enough to store our SQLite benchmarks programs and results.
\sys have a lightweight runtime, with a reduced memory footprint in the enclave, since the executable binary loaded into the enclave is only SQLite and the benchmarks.
Besides, \sys does not need an image file as it relies on the host file system, keeping its content secure thanks to IPFS.
When loaded in RAM (last lines in \Cref{tbl:cost-factors}b), the variants occupy different amounts of memory.
Native and Wasm variants store the database records in the process address space (no enclaves).
\sys and SGX-LKL store records inside their enclaves, consuming less memory outside.
The enclave sizes were configured to be just big enough to store \qty{175}{\kilo\nothing} records.

Finally, \Cref{fig:execution-time} depicts the overhead incurred by the introduction of SGX in the breakdown of the micro-benchmarks using an in-file database.
In particular, it compares the SGX hardware mode, where the SGX memory protection is enabled and the software mode, where the SGX protection is emulated.
The normalised run time is the median of the queries' execution time, from \qty{1}{\kilo\nothing} to \qty{175}{\kilo\nothing} records compared to \sys in hardware mode.
While the insertion and sequential reading time follow a similar trend, the performance of SGX-LKL in hardware mode for the random reading suffers from a slowdown.
Since SGX-LKL in software mode does not encounter this issue, the performance loss is assignable to Intel SGX.

\subsection{SQLite profiling and optimised IPFS}\label{sec:profiling}

We conducted an in-depth profiling of SQLite primitives, with particular emphasis on examining the overheads from IPFS, wherein we observed the most significant slowdowns.
The findings of this analysis have led us to the proposal of minor changes to the SGX SDK, which, remarkably, speed up the handling of protected files by a factor of \randReadingRatio$\times$.
Analogous to the previous analysis, we executed this profiling on the preliminary version of \twine (prior to upstreaming).

We instrumented and profiled IPFS, which is composed of two modules: one is statically linked with the enclave's trusted code, and the other is statically linked with the untrusted binary that launches the enclave.
We broke down these two modules into distinct components, such as cryptography, node management, and APIs for trusted and untrusted environments.
Additionally, manual instrumentation of the Wasm runtime was undertaken to profile each implemented WASI function associated with the file system.
The profiling results exclude the enclave's execution time for current time retrieval, which averages \qty{4}{\milli\second}, as its repeated usage could lead to unexpected outcomes.

We identified the primary performance contributors for random reading as follows:
\emph{(1)}~erasing memory (\texttt{memset}),
\emph{(2)}~invoking untrusted functions in the SGX SDK using \texttt{OCALL} and calling POSIX functions,
\emph{(3)}~reading the database entries, and
\emph{(4)}~internal SQLite operations (\ie cache management).
\Cref{fig:profiling} shows the costs of such operations while randomly reading the records.
Within the IPFS bar of the breakdown plot, it is evident that \memsetRatio\% of the time is allocated to clearing the enclave's memory, \ocallRatio\% to transitioning between trusted and untrusted environments (for the retrieval of the file's content), \otherOperationsRatio\% for reading operation, and a mere \sqliteRatio\% dedicated to SQLite functionality.

Internally, IPFS manages the content of a protected file by partitioning it into \emph{nodes}.
Each node corresponds to a data block that is subject to encryption or decryption.
These nodes are retained within a least recently used (LRU) cache, with each node comprising two \qty{4}{\kibi\byte} buffers designated for ciphertext and plaintext storage.
Upon a node is added to the cache, the entire data structure containing its metadata is cleared.
Given that an SGX memory page is \qty{4}{\kibi\byte} in size~\cite{cryptoeprint:2016:086}, a minimum of two pages must be cleared, in addition to the metadata encompassed within the structure, such as node identifiers and various flags.
Conversely, when a node is removed from the cache, the plaintext buffer is erased as well, corresponding to at least one SGX memory page.

Initialising structure data members is standard practice in C++ to avoid indeterminate values.
However, this practice may have a significant performance cost in SGX enclaves.
When adding nodes, functions first set specific fields before clearing the node structure.
After that, the ciphertext is transferred from the application's untrusted section into a designated buffer and decrypted into the plaintext buffer.
Given this workflow, the sole requirement for initialising class data members involves assigning default values to the unassigned fields.
We propose eliminating the clearing operations while assigning the remaining fields to zero.
Thus, we preserve the original behaviour of the code, while sparing the valuable time to clear the memory of the structure, which is overwritten in any case.
Similarly, upon a node is dropped from the cache, the plaintext buffer is cleared before releasing the node (\ie using C++'s \texttt{delete}).
Although this practice is beneficial for purging the memory of confidential values when no longer needed, we assume that SGX provides protection for the enclave's memory.
Given our threat model, no adversary can access this information, even if sensitive values remain in the SGX memory pages.
For this reason, we also propose to remove the clearing operation for the plaintext in the discarded nodes.

Finally, we examine the time spent on reading the file content. 
The function responsible for this task issues an \texttt{OCALL}, crossing the secure enclave boundary to read the content of the database file.
Our profiling measurements reveal that although untrusted POSIX calls are fast, a bottleneck exists in the code generated by the SGX tool \texttt{edger8r}, which interfaces the untrusted part of the application with the enclave. 
The \texttt{edger8r} tool eases the development of SGX enclaves by generating edge routines to interface the untrusted application and the enclave and simplifying the process of issuing \texttt{ECALL}s and \texttt{OCALL}s.
The edge functions responsible for reading files outside the enclave specify that the buffer containing the data must be copied from the untrusted application into the enclave's secure memory.
IPFS decrypts the data after issuing the \texttt{OCALL} and stores the plaintext into a buffer of the node structure.
Our profiling suggests that \copyRatio\% of the time is dedicated to completing this ciphertext copy from the untrusted application.
We propose eliminating this copy to the enclave altogether, opting instead to supply a pointer to the buffer located in the untrusted memory to the enclave, from which the library can directly decrypt.
In this revised implementation, adversaries may attempt timing attacks to alter the ciphertext between data authentication and decryption, as the authenticated mode of operation for AES-GCM is \emph{encrypt-then-MAC}.
We suggest using an alternate encryption cipher, such as AES-CCM~\cite{aesccm}, which computes the MAC from plaintext instead (\emph{MAC-then-encrypt}).
Intel's SGX SDK cryptography libraries already include this cipher.
With AES-CCM, the authentication is verified based on data stored within the enclave.
The cost of decrypting a block that fails authentication is minimal compared to a systematic buffer copy.
Moreover, it remains an infrequent event in legitimate use cases.

The performance gains of our optimised IPFS can be seen in \Cref{fig:profiling} for random reading queries with \qty{175}{\kilo\nothing} records.
The time previously devoted to clearing the memory has now been entirely eliminated, and the file reading operations now account for only \copyInvertedRatio\% of the initial execution time.
Compared to Intel's version, insertion achieves a \insertRatio$\times$ speedup and \seqReadingRatio$\times$ for sequential reading. 
Ultimately, we have achieved a \randReadingRatio$\times$ speedup for random reading.

\subsection{Macro-benchmark: \sys for Credit Scoring}\label{sec:xm}
We conclude our experimental evaluation by estimating the impact of \twine{} on the credit scoring application (see \S\ref{sec:usecases}).
Our goal is to evaluate the overhead of \twine{} for time-sensitive functionalities impacting business operations.
In particular, we focused on the \emph{Private Puller} and \emph{Aggregator} components, measuring the duration required to pull clients' data from Exchanges and to compute portfolio aggregations, respectively.
In both instances, performance is critical, as producing outdated data may impair data accuracy and subsequently undermine \xm{}'s credibility.

\begin{figure}
    \centering
    \includegraphics{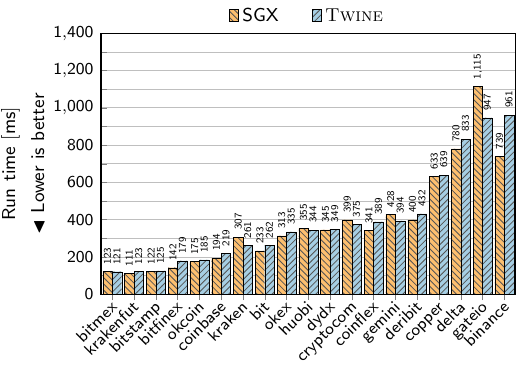}
    \caption{Pulling time of \xm{} \emph{Private Pullers} using SGX and \sys.}
    \label{fig:ppuller}
\end{figure}

The initial experiment entailed measuring the time to complete each request from single-thread clients.
We target a variety of Exchange venues by querying their API endpoints which return JSON data with a maximum size of \qty{8}{\kibi\byte}, containing clients' positions (\ie an individual's ownership in a financial asset, reflecting their investment and potential for gains or losses).
The baseline relies on an extended version of the \texttt{cpp-httplib} library~\cite{cpphttplib}, which uses SGX-WolfSSL~\cite{sgxwolfssl} to carry out TLS cryptographic functions, adopted by \xm{} in production.
On the other hand, \sys uses WolfSSL and Mongoose.
Figure~\ref{fig:ppuller} reports the results of the \emph{Private Puller} evaluation.
It is apparent that the two solutions exhibit comparable behaviour.
For some Exchanges, \twine{} is slower (\eg up to $30\%$ for Binance), while for others, it yielded better pulling time (\eg up to $17\%$ for GateIO).
This suggests that the performance is similar and subject to variations attributable to the server-side processing.

Next, we measured the duration required to complete a single aggregation round of all clients' entries (\ie a total of 522 in April 2023) available in the \xm{} production environment. 
Utilising nine parallel aggregators, we obtained the following results:
\begin{alignat*}{2}
    t(agg)_{without\_\twine}&=3\text{m}23\text{s}\\
    t(agg)_{with\_\twine}&=4\text{m}51\text{s}
\end{alignat*}
The runtime overhead using \twine{} is considered acceptable for \xm{}, which can effortlessly scale its computing units, albeit at the expense of increased infrastructural costs.
Upon further profiling, we discovered that the primary source of delay is attributable to the JSON library managing large data chunks.
\xm{} uses the \emph{nlohmann} library~\cite{Lohmann}, which is characterised by substantial memory consumption and consequently suboptimal performance when executed within \twine{}.
In future, we plan to adopt more memory-optimised JSON libraries to mitigate this overhead further.

\section{Security Analysis}\label{sec:secu}
In this section, we analyse how \twine contributes to the security of the requirements specified in \S\ref{sec:req}.
Additionally, we compare the security standpoint of \twine to some state-of-the-art solutions, namely \acctee and SGX-LKL.

\paratitle{Security of R1}The two-way sandbox in \twine offers security through two mechanisms: \emph{(1)} Intel SGX, which protects against the system tampering with \twine and the Wasm application, and
\emph{(2)}, a Wasm sandbox to enforce memory safety while requiring the hosted application to rely on WASI for any interactions with the untrusted OS.
These dual mechanisms allow both the application and infrastructure providers to confirm the integrity of the sandbox, which neither party cannot alter.
For comparison, \acctee also offers a two-way sandbox but lacks WASI support, thereby limiting the application's ability to access OS services and constraining the portability of legacy applications.
SGX-LKL, conversely, does not offer a two-way sandbox but enables the execution of legacy applications through its ad-hoc variant of libc.

We observe that AOT-compiled code can bypass the Wasm sandbox if not compiled in a secure environment.
Typically, the compilation of Wasm bytecode into assembly code ensures that memory access and function calls stay confined to the sandbox.
However, malicious actors could craft assembly code that executes unauthorised operations in \twine, like directly invoking an OCALL function.
To counter this, we suggest three mitigations: \emph{(1)} use JIT compilation to ensure secure code compilation within \twine, albeit at a performance cost, \emph{(2)} establish a separate, secure enclave solely for compilation that communicates with \twine, or \emph{(3)} coordinate with the Wasm application and infrastructure owners to validate the hash of the AOT-compiled assembly against the loaded code in \twine, provided that both parties having prior knowledge of the Wasm bytecode.
The first option is straightforward and already supported in \twine (as it is based on WAMR), while the latter two are more challenging to tackle.

\paratitle{Security of R2 and R4}In \twine, hosted applications can leverage two pre-compiled Wasm components for secure communication: a lightweight TLS library and an HTTPS library for establishing secure communication channels.
These components enable the creation of TLS-termination endpoints within the enclave, ensuring both data confidentiality and integrity during transmission.
Furthermore, \twine may extend Wasm's capability-based security model to permit application-level protocols exclusively like HTTPS, mitigating the risk of data exfiltration through insecure communication channels.
In contrast, SGX-LKL provides encrypted channels exclusively between trusted enclaves or trusted parties using Wireguard as the VPN solution.
However, this approach has limitations: both endpoints must be configured with Wireguard, and including the TCP/IP stack in the TCB contributes to its increased size.
Outside this trusted network, hosted applications must provide their own TLS implementation.
SGX-LKL's oblivious communication feature is likewise confined to its VPN network.
On the other hand, \acctee claims to support I/O calls but delegates encryption to the application or the underlying layer, which is SGX-LKL.

\paratitle{Security of R3 and R4}\ipfs is integrated with WASI for secure file operations in hosted applications, including transparent encryption and decryption.
Replacing standard libc calls with IPFS functions effectively mitigates the risk of exfiltrating sensitive data through the file system.
The enclave's sealing key serves as the basis for a symmetric key, decrypting a file's Merkle tree root node.
This root node holds essential metadata for decrypting subsequent nodes~\cite{IPFSexplained}.
Although the Merkle tree nodes are stored on the untrusted file system, decryption is limited to the specific enclave or its owner via the enclave- or owner-bound sealing key~\cite{IntelSgxSealing}.
SGX-LKL also offers file system security through the use of virtual block devices.
It creates virtual disks on an untrusted file system and uses an ad-hoc algorithm to mitigate side-channel data leaks.
While \twine does not guard against side-channel attacks, its abstraction through WASI allows the secure file system to be replaced with other state-of-the-art solutions like Obliviate~\cite{ahmad2018obliviate}.
As for \acctee, file system security is delegated to SGX-LKL, similar to its approach to network security.

\paratitle{Security of R5}\twine offers hosted applications the capability for attesting the Wasm bytecode in JIT mode, or assembly code in AOT mode, along with the runtime.
This serves two purposes: \emph{(1)} it lets the enclave owner validate the genuineness of the hardware and the integrity of the application, and \emph{(2)} it provides assurance to the infrastructure owner of the correct implementation of the \twine runtime.
Furthermore, the attestation process ensures that a specific configuration of \twine is in place, including disabling some system calls for the Wasm module.
Turning off specific system calls enhances security by reducing the attack surface, adhering to the principle of least privilege, and easing system monitoring and auditing.
By comparison, \acctee mentions attestation but lacks an API to expose the evidence to the Wasm applications for secure communication. 
SGX-LKL offers remote attestation by sharing the hash of the virtual disks with trusted entities.
However, this approach can be challenging, especially when hosted applications store files since the attestation measurement also reflects these files.
Consequently, managing multiple virtual disks becomes necessary for maintaining known attestation measurements.

\section{Conclusion}\label{sec:concl}

The reluctance to adopt distributed architectures for sensitive applications arises from a lack of trust when outsourcing computations to remote parties.
Although this issue has been extensively studied in the context of TEEs, such solutions introduce non-trivial drawbacks and constraints, including limited programming language support, restrictions on system calls, and enforced programming paradigms.
In this paper, we propose an approach for executing unmodified programs in WebAssembly (Wasm)---a target binary format for applications written in LLVM-supported languages, such as C, C++, Rust, Go, and Swift---within lightweight TEEs that can be easily deployed across client and edge computers.
\twine is our trusted runtime that supports the execution of unmodified Wasm binaries within SGX enclaves.
Wasm offers several advantages, including speed, versatility, and abstraction of the complexity associated with developing applications tailored for specific TEEs. 
Furthermore, we provide an adaptation layer between the standard WebAssembly system interface (WASI) used by applications and the underlying OS, translating WASI operations into equivalent native system calls or functions from secure libraries specifically designed for SGX enclaves.
Consequently, trusted applications can seamlessly interact with encrypted files and secure network connections via TLS and HTTPS.
Our comprehensive evaluation demonstrates performance comparable to other state-of-the-art approaches while offering robust security guarantees and full compatibility with standard Wasm applications.
Finally, \twine is freely available as open-source software and has been merged into the original WAMR runtime.

\ifCLASSOPTIONcompsoc
\paratitle{Acknowledgments}
\else
\paratitle{Acknowledgment}
\fi
This publication incorporates results from the VEDLIoT project, which received funding from the European Union's Horizon 2020 research and innovation programme under grant agreement No 957197.
We extend our sincere appreciation to \xmi for their sponsorship and contributions to this research paper. \ifCLASSOPTIONcaptionsoff
  \newpage
\fi

\printbibliography

\vspace{100mm}
\begin{IEEEbiography}[{\includegraphics[width=1in,height=1.25in,clip,keepaspectratio]{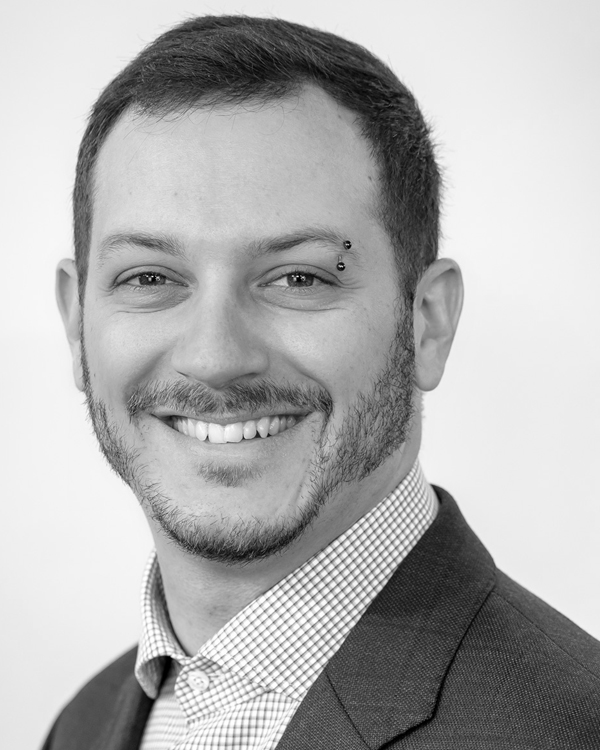}}]{Jämes Ménétrey} received the B.Sc. degree in Computer Science and the M.Sc. degree in Engineering from the University of Applied Sciences and Arts Western Switzerland (HES-SO), in 2018 and 2020, respectively.
Since 2020, he is a Ph.D. student at the University of Neuch\^atel (Switzerland).
His research is focused on efficient dependable computing and software security using emerging technologies, such as WebAssembly and trusted execution environments.
\end{IEEEbiography}

\vspace{-17mm}
\begin{IEEEbiography}[{\includegraphics[width=1in,height=1.25in,clip,keepaspectratio]{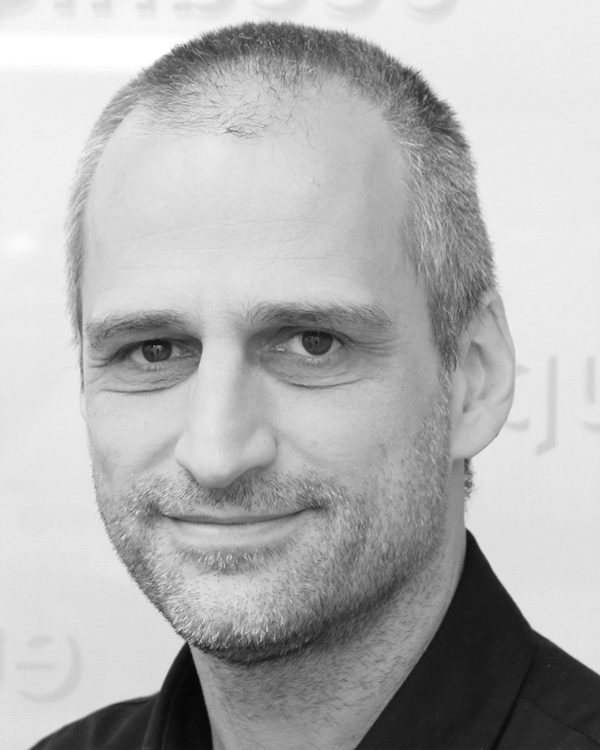}}]{Marcelo Pasin} received his PhD from the Grenoble Institute of Technology (INPG), and is a member of IEEE, ACM and SBC (Brazil).
Researcher in the University of Neuchâtel (Switzerland) and an associate professor in the Engineering School Arc of the University of Applied Sciences and Arts Western Switzerland (HES-SO), he currently conducts research on distributed execution environments for the cloud continuum.
\end{IEEEbiography}

\vspace{-17mm}
\begin{IEEEbiography}[{\includegraphics[width=1in,height=1.25in,clip,keepaspectratio]{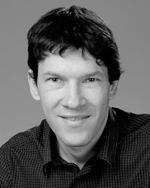}}]{Pascal Felber} received his M.Sc. and Ph.D. degrees in Computer Science from the Swiss Federal Institute of Technology (EPFL). 
He has then worked at Oracle Corporation and Bell-Labs in the USA, and at Institut EURECOM in France.
Since 2004, he is a Professor of Computer Science at the University of Neuch\^atel, Switzerland, working in the field of dependable, distributed, and concurrent systems. 
He has published over 200 research papers in various journals and conferences.
\end{IEEEbiography}

\vspace{-17mm}
\begin{IEEEbiography}[{\includegraphics[width=1in,height=1.25in,clip,keepaspectratio]{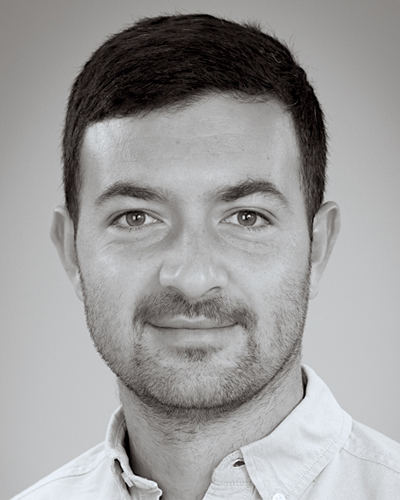}}]{Valerio Schiavoni}, PhD, Member of IEEE, is Ma\^itre-Assistant (Lecturer) at the University of Neuch\^atel, Switzerland.
His research interests lie at the intersection of systems broadly conceived, security, and data management.
\end{IEEEbiography}
 
\vspace{-14mm}
\begin{IEEEbiography}[{\includegraphics[width=1in,height=1.25in,clip,keepaspectratio]{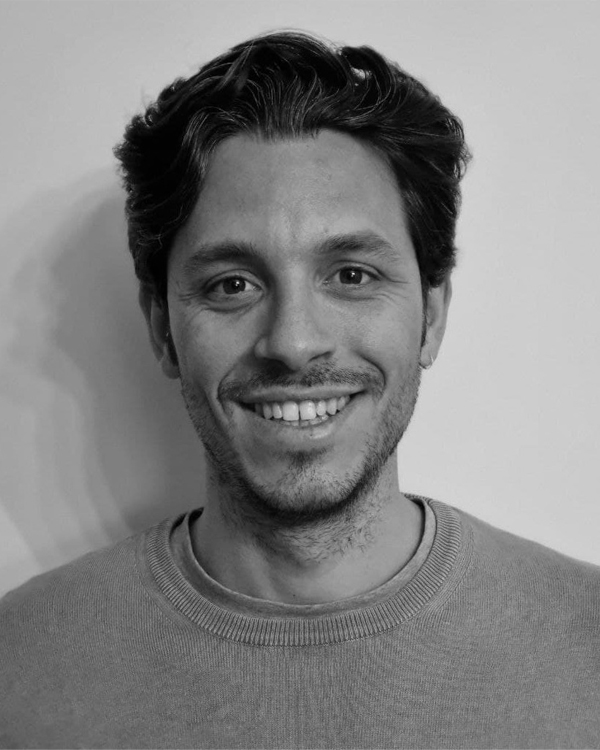}}]{Giovanni Mazzeo}, PhD, is Assistant Professor at the Department of Engineering of the University of Naples "Parthenope”. 
His research activity mainly focuses on security and dependability of computer systems. 
He is actively involved in European projects on IT security.
\end{IEEEbiography}

\vspace{-14mm}
\begin{IEEEbiography}[{\includegraphics[width=1in,height=1.25in,clip,keepaspectratio]{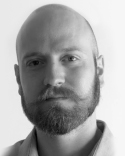}}]{Arne Hollum} is the Co-Founder and CTO of Credora, an end-to-end lending solution facilitating credit by validating real-time risk metrics in a zero-knowledge environment. 
Arne started his career building quantitative trading and machine learning systems across fixed-income products at ING before co-founding Credora in 2019.
\end{IEEEbiography}

\vspace{-14mm}
\begin{IEEEbiography}[{\includegraphics[width=1in,height=1.25in,clip,keepaspectratio]{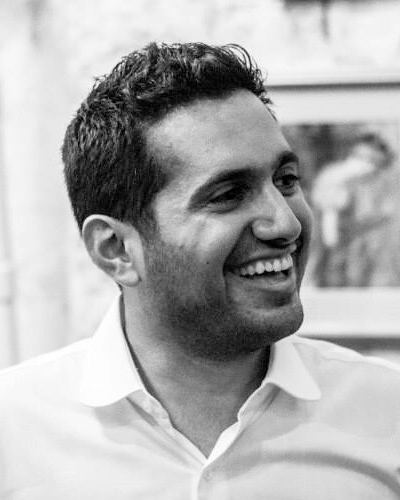}}]{Darshan~Vaydia} is the Co-Founder and CEO of Credora, the private credit oracle. 
	Credora is an end-to-end lending solution facilitating credit by validating real-time risk metrics in a privacy preserving environment.
	Darshan started his career as an options trader at UBS \& Mako Global and ran one of the early crypto options market makers before co-founding Credora in 2019.
\end{IEEEbiography}
\end{document}